\documentclass[12pt, oneside]{amsart}
\usepackage{amssymb}
\usepackage{graphicx, bbm}
\usepackage{url,tabularx,array,geometry,appendix, rotating}
\usepackage{multirow}
\usepackage{amsfonts}
\usepackage{color}
\usepackage{fullpage}

\def\bSig\mathbf{\Sigma}

\newcommand{\bs}{\boldsymbol{x}}

\newcommand{\bsK}{\boldsymbol{K}}
\newcommand{\bst}{\boldsymbol{\theta}}

      \makeatletter
      \def\@setcopyright{}
      \def\serieslogo@{}
      \makeatother
   
\begin{document}



   \author{Silvia Montagna}
   \address{Dept. of Statistics\\
		University of Warwick \\
		Coventry CV4 7AL \\
		UK}
   \email{S.Montagna@warwick.ac.uk}


   \author{Surya T.~Tokdar}
   \address{Dept. of Statistical Science\\
			Duke University \\
			Box 90251 \\
			Durham, NC, USA}
   \email{st118@stat.duke.edu}
   

   \title[Computer emulation with non-stationary GP's]{Computer emulation with non-stationary Gaussian processes}


\begin{abstract}
Gaussian process (GP) models are widely used to emulate propagation uncertainty in computer experiments. GP emulation sits comfortably within an analytically tractable Bayesian framework. Apart from propagating uncertainty of the input variables, a GP emulator trained on finitely many runs of the experiment also offers error bars for response surface estimates at unseen input values. This helps select future input values where the experiment should be run to minimise the uncertainty in the response
surface estimation. However, traditional GP emulators use stationary covariance functions, which perform poorly and lead to sub-optimal selection of future input points when the response surface has sharp local features, such as a jump discontinuity or an isolated tall peak. We propose an easily implemented non-stationary GP emulator, based on two stationary GPs, one nested into the other, and demonstrate its superior ability in handling local features and selecting future input points from the boundaries of such features.
\end{abstract}


   \keywords{Bayesian inference; Computer emulation; Non-stationary Gaussian process; Sequential design; Particle learning; Uncertainty quantification.}



   \date{\today}


   \maketitle
\section{Introduction}
\begin{figure}[t]
	\centering
	\includegraphics[scale = 0.70]{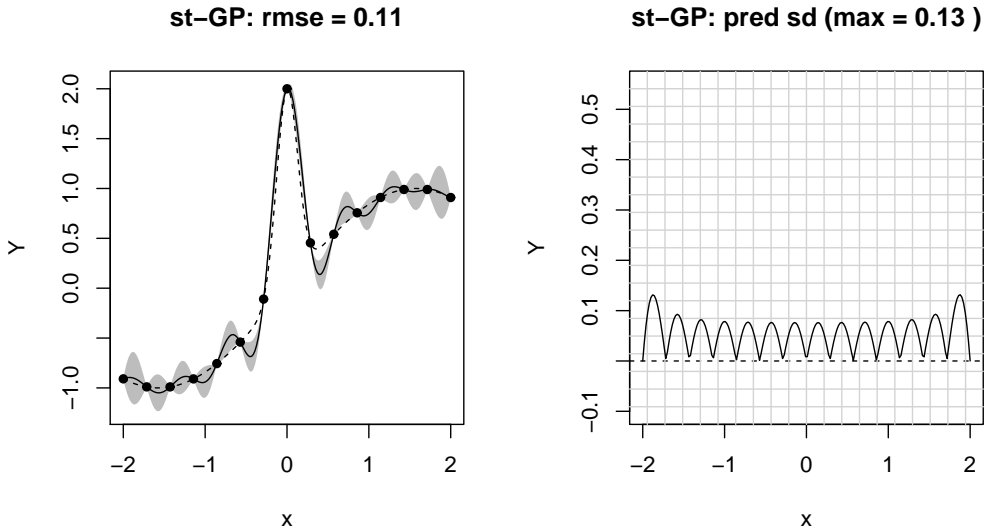}
	 \vspace{-.6cm}
	 \caption{Plot of (true) function $f(x) = \sin(x) + 2\exp(-30x^2), x \in [-2, 2]$ (dashed line). The black dots represent observed data at 15 equally-spaced values of $x$. Left panel: the solid line is the point predictor of $f$, or conditional mean, obtained from a stationary GP emulator fitted to the data. Shaded areas represent the error bars. Right panel: standard deviation evaluated at 200 predictive locations.}
	\label{peak}
	\end{figure}

Large scale computer simulation is widely used in modern scientific research to investigate physical phenomena that are too expensive or impossible to replicate directly \cite{Schade1999, Fan2009, Textor2009}. Most simulators depend on a handful of tuning parameters and initial conditions, referred to as the input arguments. Often interest focuses on quantifying how uncertainty in the input arguments propagates through the simulator and produce a distribution function over one or many outputs of interest. In this paper we consider only deterministic simulators which when run on the same input twice will produce identical output values. \\ 
\indent Quantifying uncertainty propagation will require several runs of a simulator at different input points to learn the input-output map ${Y} = f(\bs)$ accurately over the entire input space. However, computer simulations are very time-consuming, thus running a simulator over a dense grid of input points could be prohibitively expensive. On the other hand, running a simulator over a sparse design chosen in advance may result in insufficient information in vast parts of the input space. Consequently, there is considerable interest in estimating a slow computer simulator with a fast statistical ``emulator''  \cite{Sacks1989, Kennedy2001, Santner2003}. The emulator is fitted to input-output data $\{\bs^t, f^t\}$, where $f^t = \{f(\bs_1), \dots, f(\bs_t)\}$ is obtained from a few preliminary runs of the simulator on design $\bs^t = \{\bs_1, \dots, \bs_t\}$, and the fitted model is then used for prediction of $f$ at input configurations not included in $\bs^t$ \cite{Sacks1989, Busby2009}.  \\
\indent For Bayesian emulation, a common practice is to assign $f$ a Gaussian process prior \cite{Sacks1989, Currin1991, Schmidt2000}. Gaussian process (GP) emulation is appealing due to its mathematical tractability and ability to incorporate a wide range of smoothness assumptions. The conditional posterior distribution of $f$ at future inputs, given data $\{\bs^t, {f}^t\}$ and process hyperparameters, remains a GP distribution. The posterior mean of $f(\bs)$ gives a statistical estimate or surrogate for the simulator output at a new input $\bs$, whereas the posterior variance at $f(\bs)$ quantifies how well the simulator has been learned at and around $\bs$. The latter is a particularly attractive feature of GP emulation as it provides a model-based assessment of the emulator's accuracy and could be used to actively learn an optimal sequence of input points on which the simulator needs to be run to minimise the uncertainty in posterior surface estimation. \\
\indent Research on computer emulation has largely focused on stationary GP models \cite{Sacks1989, Kennedy2001}. Stationary GPs regard the similarity between $f(\bs)$ and $f(\bs + \boldsymbol{h})$ as a decaying function in $\boldsymbol{h}$ only, known up to global smoothness and decay parameters. This is a strong prior assumption that is not easily washed away by data and may lead to unrealistic emulation for many physical phenomena. In practice, stationary GP emulators run into difficulties when the shape of $f$ has sharp localised features, e.g. abrupt discontinuities or tall peaks, and lead to poor point predictions and selection of future inputs. A simple example is illustrated in the left panel of Figure \ref{peak}. Three aspects emerge: (i) the discovery of a tall peak in the middle has a rippling effect and creates large oscillations of the predictive mean curve over a large part of the input space, a phenomenon often called ``spline tension" effect in the predictor form; (ii) prediction seems overconfident around the peak, where the error bars are too narrow to capture the high variability around $x = 0$; instead, (iii) prediction intervals are quite large where abrupt changes in the function values are {\textit{not}} observed, and $f$ is relatively more well-behaved. A sequential design strategy based on uncertainty quantified by the prediction standard deviation would favour the selection of a new input from the whole $x$ domain, with the only exception of the tall peak (right panel in Figure \ref{peak}). Thus, stationary GPs favour the selection of new points in unexplored regions of the input space ({\textit{exploration}}), but tend to neglect regions that are deemed important based on the current estimate of $f$ ({\textit{exploitation}}). \\
\indent Extrinsic diagnostics is often used to assess the adequacy of a GP emulator as surrogate for the simulator \cite{Bayarri2007,Bastos2009}. For example, one can examine the leave-one-out cross validated (CV) standardised residuals to quantify the emulator's uncertainty. Either too large or very small CV standardised residuals (as compared to a $N(0,1)$ or a $t_\nu$) at some validating points indicate that the emulator is poorly estimating the predictive uncertainty. Outliers of this kind denote a local fitting problem, which could be improved upon by adding new points in the vicinity. Thus, CV examines the local behaviour of $f$, and flags those sub-regions where the simulator has more variations. Therefore, CV leans toward an {\textit{exploitation}}-driven sequential design. Although CV is often combined with a stationary GP to better address sequential design, it is difficult to reconcile the {\textit{exploration}}-driven predictive variance of a stationary GP with the {\textit{exploitation}}-driven flagging of CV, and any combination is ad-hoc. Also, the model remains misspecified: a stationary model is used for a response which is often intrinsically not so \cite{Busby2009}. \\
\indent Several approaches proposing non-stationary GP models can be found in the literature. In the context of computer emulation, \cite{Gramacy2008} propose the Bayesian treed GP model (TGP), which applies independent stationary GPs to subregions of the input space determined by data-driven recursive partitioning parallel to the coordinate axes. Because of the parallel partitioning, TGP adapts well to surfaces having {\textit{rectangular}} local features (``axes-aligned" non-stationarity). However, it may run into difficulties when the nature of the non-stationarity is more general. Also, TGP's hard partitioning of the input space prevents borrowing of information across partitions and enforces discontinuity on the estimated response surface. \cite{Ba2012} decompose $f$ into the sum of two stationary GPs, the first capturing the smooth global trend and the second modelling local details. Other approaches in the context of GP regression include \cite{Sampson1992, Schmidt2000, Paciorek2004}. This literature makes it clear that the main challenges in non-stationary GP modelling are to keep the number of hyperparameters under control to facilitate efficient learning from limited data while allowing for non-stationary features of various geometric shapes and at the same time not to enforce non-stationarity when not needed.\\ 
\indent This paper is designed to serve two purposes. First, to introduce a non-stationary GP emulator which is adaptable to local features of many kinds of shape. Second, to use our emulator for online learning of an optimal sequence of design points. Non-stationarity is achieved by augmenting the input space with one extra latent input which we infer from the data. The latent input can flag regions of the input space characterised by abrupt changes of the function values and help correct for inadequacies in the fit. For sequential design, it is absolutely crucial to have trustworthy judgement of uncertainty of the current estimate of $f$ to concentrate efforts only on where needed. Sections \ref{sequential higher} and \ref{LGBB} show results from various synthetic and real experiments where a sequential version of our emulator outperforms similar sequential adaptations of existing GP emulators, when performance is measured by the number of simulator runs needed to achieve a certain accuracy. \\
\indent The proposed method is also attractive from an operational point of view. Both the latent input dimension and the response function (of the original plus the latent inputs) are individually modelled as stationary GPs controlled by a small number of hyperparameters that can be efficiently learned with sequential Monte Carlo (MC) computing leveraging on conjugacy properties of GP. Sequential MC computing seamlessly blends with active learning of the sequential design, as opposed to Markov chain sampling based non-stationary GP emulators whose sequential adaptation requires re-running the whole Markov chain sampler at every iteration. \\
\indent The remainder of the paper is outlined as follows. Section \ref{GPemulation} begins with an overview of GP emulation and stationarity and then introduces our non-stationary GP emulator. Section \ref{Implementation} presents a fast sequential design algorithm for GP emulation. Section \ref{studies} examines the performance of different emulators in quantifying uncertainty through one-dimensional numerical examples. In Section \ref{sequential higher}, we investigate sequential design via higher-dimensional examples. Section \ref{LGBB} presents a real data application. Conclusions are reported in Section \ref{conclusions}.

\section{Gaussian process emulators}
\label{GPemulation}

\subsection{GP emulation and stationarity}
\label{GPemulationstationary}
The canonical emulator used for the design and analysis of computer ex\-pe\-ri\-ments is the GP. Specifically, for any finite collection of inputs $(\bs_1, \dots, \bs_t)^\top$, the joint distribution of $(f(\bs_1), \dots, f(\bs_t))^\top$ is multivariate Gaussian with mean $\mathbbm{E}[f(\bs)] = \mu(\bs)$ and positive definite covariance matrix $\mathbbm{C}\text{ov}[f(\bs), f(\bs^\prime)] = C_{\bst}(\bs, \bs^\prime) = \sigma^2K_{\bst}(\bs, \bs^\prime)$ parameterised by $\bst$. Note that we can write the GP emulator as 
	\begin{equation}
	\label{firstGP}
	f(\bs) = \mu(\bs) + \epsilon(\bs; \boldsymbol{\theta}), 
	\end{equation}
where $\epsilon(\bs; \boldsymbol{\theta})$ is a zero-mean GP with covariance function $C_{\bst}(\cdot, \cdot)$. To simplify the notations, we shall drop the $\bst$ subscript to the covariance and correlation functions hereafter.  \\ 
\indent The representation of $f$ as a Gaussian vector makes the computation conceptually straightforward. The conditional distribution of $f$ at a new input $\tilde{\bs}$, given data $\{\bs, f(\bs)\}_{1:t} \equiv \{\boldsymbol{X}, \boldsymbol{F}\}$ and parameters $\boldsymbol{\theta}$, is also Gaussian with mean 

$$\hat{f}(\tilde{\bs}) = \mathbbm{E}[f(\tilde{\bs}) \vert \{\bs, f(\bs)\}_{1:t}, \boldsymbol{\theta}] = \mu(\tilde{\bs}) + k^\top(\tilde{\bs})\bsK^{-1} (\boldsymbol{F} - \mu(\boldsymbol{X}))$$ 

\noindent and variance

$$\hat{\sigma}^2(\tilde{\bs}) = \mathbbm{V}[f(\tilde{\bs}) \vert \{\bs, f(\bs)\}_{1:t}, \boldsymbol{\theta}] = \sigma^2 \{K(\tilde{\bs}, \tilde{\bs}) - k^\top(\tilde{\bs})\bsK^{-1}k(\tilde{\bs})\}$$

\noindent where $k^\top(\tilde{\bs})$ is the $t-$vector whose $i$-th component is $K(\tilde{\bs}, \bs_i), i = 1, \dots, t$, and $\bsK$ is the $t\times t$ correlation matrix with $i,j$ element $K(\bs_i, \bs_j)$. \\ 
\indent The mean field $\mu(\bs)$ in (\ref{firstGP}) is typically given the linear model structure $\mu(\bs) = h(\bs)^\top\boldsymbol{\beta}$, where $\boldsymbol{\beta}$ is a vector of unknown parameters. Although $h(\cdot)$ may be any function on the input space $\mathcal{X}$, we adopt a linear mean in the inputs, $h(\bs) = [1, x_{1}, \dots, x_{p}]^\top$. This seems to be a natural choice with little prior information about the input-output relationship and helps to control overfitting. The correlation function is crucial in GP modelling;  it is through $K(\bs, \bs^\prime)$ that we express a belief about how similar $f(\bs)$ and $f(\bs^\prime)$ should be if $\bs$ and $\bs^\prime$ were close in $\mathcal{X}$, thereby we express a belief about the smoothness of $f$. Although different formulations are possible \cite{Rougier2009}, in this work we focus on the power family and use the separable power correlation function 
	\begin{equation} 
	K(\bs, \bs^\prime) = e^{-\sum_{l=1}^p \phi_l(x_{l} - x^\prime_{l})^{p_0}},
	\label{stationary} \end{equation}
which is a standard choice in modelling computer experiments \cite{Santner2003}. We fix $p_0 = 2$ (product-Gaussian correlation) and infer the correlation range parameters $\{\phi_l\}_{l = 1}^p$ as part of our estimation procedure. Thus, the correlation is only function of $\bs - \bs^\prime$ (stationarity) and a set of roughness (unknown) parameters. \cite{Van2009} show that the squared-exponential kernel (\ref{stationary}) can optimally adapt to any smoothness level. \cite{Bhattacharya2011} develop a class of priors for the correlation range parameters which leads to minimax adaptive rates of posterior concentration. \\
\indent We embed our approach in a Bayesian framework and proceed by specifying prior distributions for the model parameters. Hereafter, we use an improper uniform prior $\boldsymbol{\beta} \propto 1$ as conventional representation of weak prior information about $\boldsymbol{\beta}$, an inverse-gamma (IG) prior for the scale, $\sigma^2 \sim \text{IG}(a/2, b/2)$, and a log-normal prior for the correlation parameters, $\phi_l \sim \log \text{N}(\mu_\phi, \nu_\phi)$, but other formulations are possible \cite{Gramacy2008}. The posterior predictive distribution of $f$ at a new input $\tilde{\bs}$, conditioned on $\bsK$ and data observed up to time $t$ and marginalised with respect to $\{\boldsymbol{\beta}$, $\sigma^2\}$, is a Student-$t$ distribution with $\hat{\nu} = t - p - 1$ degrees of freedom, mean 
	\begin{equation}
	\hat{f}(\tilde{\bs} \vert \{\bs,f(\bs)\}_{1:t}, \bsK) = h(\tilde{\bs})^\top \tilde{\boldsymbol{\beta}} + k^\top(\tilde{\bs})\bsK^{-1}(\boldsymbol{F} - \boldsymbol{H}^t\tilde{\boldsymbol{\beta}}), \label{fhat}
	\end{equation}
and variance 
	\begin{equation}
	\hat{\sigma}^2(\tilde{\bs} \vert \{\bs, f(\bs)\}_{1:t}, \bsK) = \frac{(b + \Phi) [K(\tilde{\bs}, \tilde{\bs}) - k^\top(\tilde{\bs})\bsK^{-1}k(\tilde{\bs})]}{a + \hat{\nu}}, \label{sigmahat}
	\end{equation}
where $\boldsymbol{H}^t$ is the $t \times (p+1)$ matrix which contains $h(\bs_i)^\top$ in its rows and
	\begin{align*}
	\Phi &= \boldsymbol{F}^\top \bsK^{-1}\boldsymbol{F} - \tilde{\boldsymbol{\beta}}^\top \Psi^{-1} \tilde{\boldsymbol{\beta}} \\
	\tilde{\boldsymbol{\beta}} &= \Psi(\boldsymbol{H}^t \bsK^{-1} \boldsymbol{F}) \\
	\Psi &= (\boldsymbol{H}^t \bsK^{-1} \boldsymbol{H}^t)^{-1}.
	\end{align*}
The availability in closed form of the marginalised predictive distribution is crucial for the sequential design algorithm implementing our non-stationary GP emulator (\S \ref{Implementation}).  

 \subsection{Non-stationary GP through latent input augmentation}
 \noindent In computer emulation, it is not uncommon to observe functions that vary more quickly in some parts of the input space than in others \cite{Gramacy2008}. In response to concerns about the adequacy of the stationary assumption for GP emulators, we build on the concept of spatial deformation \cite{Sampson1992} and model $f(\bs)$ as
	\begin{equation}
	f(\bs) = \mu(\bs) + \epsilon([\bs, Z]; \boldsymbol{\theta}), 
	\end{equation}
where $\mu(\bs) = h(\bs)^\top\boldsymbol{\beta}$ and $\epsilon([\bs, Z]; \boldsymbol{\theta})$ is a zero-mean GP whose covariance function depends smoothly on the $p-$dimensional (known) vector of inputs, $\bs \in \mathcal{X} = \mathbb{R}^p$, a latent (unknown) input $Z$ which we infer from the data, and a handful of model parameters $\bst$. Specifically, we adopt an ``augmented'' product-Gaussian correlation form for $K = \sigma^{-2}C$:
		\begin{equation}	
		{K}(\bs_i, \bs_j) = \exp\left\{- \sum_{l = 1}^p \phi_l(x_{il} - x_{jl})^2 - \phi_{p+1}(Z_i - Z_j)^2\right\}.
		\label{Z}
		\end{equation} 
Expression (\ref{Z}) corresponds to the standard (squared-exponential) correlation function of a stationary GP (\ref{stationary}) indexed by $p+1$ inputs. \\
\indent Clearly, the problem of modelling a non-stationary simulator could be tackled differently. For example, \cite{Rougier2009} propose gathering substantial information about the simulator from experts and use it to include a large number of regressors in the prior mean field $\mu$ while retaining a stationary residual process. However, choosing the best set of regressors is non-trivial and there is no guarantee that the residual process is stationary. Also, the authors suggest using rougher (but still stationary) correlation functions like the Mat\'ern. While this could lead to a better fit compared to smoother alternatives, it would require specifying an extra smoothing parameter which is hard to infer statistically. Also, the stationary representation does not address the issue that stationary GP processes focus more on exploration rather than exploitation of the input space, thus still leading to a sub-optimal sequential design strategy in presence of local features. \cite{Paciorek2004} generalise Gibbs' construction to obtain non-stationary versions of arbitrary isotropic covariance functions. While their model provides a flexible and general framework, it is computationally demanding and not feasible in high-dimensional spaces. The latent extension of the input space guarantees positive definiteness of the covariance between observations in the original space and enhances an intuitive interpretation of the problem. When thinking of emulation of computer models that are characterised by sharp local features, the extra input could tear apart regions of the input space that are separated by abrupt changes of the function values. The correlation between points at and about a localised feature is weakened since the corresponding distance has been stretched by the latent coordinate.\\
\indent To this point we have not made any assumptions about the latent input $Z$. In the following, we model $Z$ as a continuous function of the inputs, $Z_i = g(\bs_i)\in \mathbb{R}$, using a stationary GP 
	\begin{align}
	\label{Zprocess}
	g \mid  \bst &\sim \text{GP}(0, \tilde{K}), \qquad \text{with} \\
	\tilde{K}(\bs_i, \bs_j) &= \exp\left\{- \sum_{l = 1}^p \tilde{\phi}_l (x_{il} - x_{jl})^2\right\},
	\end{align} 
	where the scale parameter is fixed to 1. \\
\indent To summarise, our formulation relies on two stationary GPs, one for the function of interest and one for the latent input 
			\begin{equation}
			f \vert \bst \sim \text{GP}(\mu, C), \qquad \text{and} \qquad g \vert \bst \sim \text{GP}(0, \tilde{K}),
			\label{construction}
			\end{equation}
where vector $\bst$ collects all the parameters of both the original and latent processes, $\bst = [\boldsymbol{\beta}, \sigma^2, \{\phi_l\}_{l=1}^{p+1}, \{\tilde{\phi}_l\}_{l=1}^{p}]$. A stochastic process for $f(\bs)$ is achieved by integrating out the regression parameters $\boldsymbol{\beta}$ and $Z$, and is more adaptive than (\ref{firstGP}) to functions whose smoothness varies with the inputs because it has the capacity to have several length scales. \\	
\indent As mentioned above, \cite{Sampson1992} first pioneered an approach to the problem of nonstationarity and anisotropy in environmental datasets through a nonlinear transformation of the sampling space into a latent space with stationary and isotropic spatial structure. The mapping was done via multidimensional scaling. Further, the authors used thin-plate splines to estimate realisations of $f$ at predictive locations while keeping the estimates of the latent process fixed and without taking into account any measure of uncertainty about the mapping \cite{Schmidt2000}. The approach we propose in (\ref{construction}) is similar in flavour to the construction in \cite{Schmidt2000}, who build on \cite{Sampson1992} and implement spatial deformation via a GP prior. However, our construction differs from \cite{Schmidt2000}, where $K$ is chosen to correspond to a mixture of Gaussian correlation functions, each of which depends on the Euclidean distance between the latent inputs $Z$'s only. Also, the authors infer their deformation from an observation of a sample covariance matrix. The idea of achieving nonstationarity by latent input extension can also be found in \cite{Pfingste2006}, who present two approaches for approximate Bayesian inference in GP regression models. The first method relies on a discrete latent input and is implemented in an Markov Chain Monte Carlo (MCMC) sampling scheme, whereas the second method estimates a continuous latent mapping by evidence maximisation. We remark that \cite{Sampson1992, Schmidt2000, Paciorek2004, Pfingste2006} are not attempting to estimate a deterministic model or perform sequential design. 
 
\section{Implementation}
\label{Implementation}
We apply our adaptive non-stationary GP emulator to the sequential design of computer experiments \cite{Santner2003}: start with a parsimonious design; choose a new input $\bs$ according to some criterion derived from the emulator fit; update the emulator conditional on the new pair $\{\bs, f(\bs)\}$; and repeat until some stopping criterion is met. Sequential design is crucial to keep designs small and save on expensive runs of the simulator while guaranteeing adequate learning of the input-output map. We adopt a sequential MC technique known as particle learning (PL) to obtain a quick update of the emulator after each sequential design iteration (\S \ref{SMC}) and discuss some sequential design criteria in \S \ref{Adaptive sequential design}. An introduction to PL is beyond the scope of this paper. The unfamiliar reader can refer to \cite{Lopes2011} for an introduction and to \cite{Gramacy2011} for its application to the online updating of GP regression models. 

\subsection{PL for GP emulation}
\label{SMC}
PL provides a simulation-based approach to sequential Bayesian computation. Central to PL is the identification of {\it essential state vectors} or particles, $\{S_t^{(i)}\}_{i = 1}^N$, that are tracked sequentially, with $N$ denoting the total number of particles. These particles contain all the sufficient information about the uncertainties given the data up to time $t$ and are used to approximate the posterior distribution, $\{S_t^{(i)}\}_{i = 1}^N \sim \pi(S_t \mid \{\bs,f(\bs)\}_{1:t})$. PL provides a method to update the particles from $t$ to $t+1$. \\
\indent We start by identifying the quantities particles include. The sufficient information necessarily depends upon $[(\bs_1, f(\bs_1)), \dots, (\bs_t, f(\bs_t))]$ \footnote{To stress the dependence of $f$ to both known and latent inputs within our approach, we should write $f(\bs, Z)$. In the remainder, however, we will omit $Z$ and write $f(\bs_1), \dots, f(\bs_t)$ to simplify the notation. }, thus $\{S_t^{(i)}\}_{i = 1}^N = \{(Z_{1:t}, K_t, \tilde{K}_t)^{(i)}\}$, with $Z_{1:t} \equiv (Z_1, \dots, Z_t)^T$. The correlation functions have been indexed by $t$ to stress their dependence to data collected up to time $t$. Particles do not contain $\boldsymbol{\beta}$ nor $\sigma^2$ as these parameters can be marginalised out within our Bayesian construction \cite{Gramacy2011}. \\
\indent Suppose we start with $(f(\bs_1), \dots, f(\bs_{t_0}))^\top$ obtained from $t_0 > p + 1$ preliminary runs of the simulator at design points $(\bs_1, \dots, \bs_{t_0})^\top$. The initial design can be chosen as, e.g., a Latin Hypercube Design (LHD). Particles are initialised at time $t_0$ with a sample of the unknown parameters from their prior distributions. In addition to the priors in \S \ref{GPemulationstationary}, we also sample $\{\tilde{\phi}_l\}_{l = 1}^p$ from $\tilde{\phi}_l \sim \text{log N}(m_{\tilde{\phi}}, v_{\tilde{\phi}})$. The core components of PL for updating particles $\{S_t^{(i)}\}_{i = 1}^N$ to $\{S_{t+1}^{(i)}\}_{i = 1}^N$ are the following two steps: \\

	\begin{itemize}
	\item \textit{Resample} Generate index $\zeta \sim \text{Multinomial}(w,N)$, with
				$$w^{(i)} = \frac{\pi(f(\bs_{t+1}) | S_t^{(i)})}{\sum_{i = 1}^N \pi(f(\bs_{t+1}) | S_t^{(i)})}, \qquad i = 1,\dots, N,$$
				where $\pi(f(\bs_{t+1}) | S_t^{(i)}) = \pi(f(\bs_{t+1}) | [\bs, f(\bs)]_{1:t}, K_t^{(i)})$ denotes the probability of observing $f(\bs_{t+1})$ under a Student-$t$ distribution with $\hat{\nu} = t - p - 1$ degrees of freedom and mean and variance given by Equations (\ref{fhat})--(\ref{sigmahat}), respectively. \\
				
	\item \textit{Propagate} each resampled particle $S_t^{\zeta(i)}$ to $S_{t+1}^{(i)}$ to account for $[\bs_{t+1}, f(\bs_{t+1})]$\\
		\begin{itemize}
		\item Construct the ``propagated'' correlation fun\-cti\-on of the latent GP, which will be used to sample the latent coordinate at the new input $\bs_{t+1}$. Thus, we build $\tilde{\bsK}_{t+1}^{(i)}$ from $\tilde{\bsK}_{t}^{(i)}$ and $\tilde{k}_t^{(i)}(\bs_{t+1}) = \tilde{K}^{(i)}(\bs_{t+1}, \bs_j)$, with $j = 1, \dots, t$
			 \[
 			\tilde{\bsK}_{t+1}^{(i)} = \begin{bmatrix}
          					\tilde{\bsK}_{t}^{(i)}  & \tilde{k}_t^{(i)}(\bs_{t+1})      \\[0.3em]
      					 	\tilde{k}_t^{(i)\top}(\bs_{t+1}) & \tilde{K}^{(i)}(\bs_{t+1}, \bs_{t+1})
					  \end{bmatrix} \]
		\item Obtain $g^{(i)}(\bs_{t+1})$ from its predictive distribution $g^{(i)}(\bs_{t+1}) \mid g^{(i)}(\bs_{1:t}), \tilde{\bsK}^{(i)}_{t+1} \sim \text{N}({\mu}^{*(i)}, \tilde{K}^{*(i)})$, where mean and variance are obtained via standard kriging equations \\
	   \item Construct the ``propagated'' correlation function of $f$. We build $\bsK_{t+1}^{(i)}$ from $\bsK_{t}^{(i)}$ and $k_t^{(i)}(\bs_{t+1}) = K^{(i)}(\bs_{t+1}, \bs_j)$, $j = 1,\dots, t$, as
		 	\[
 			\bsK_{t+1}^{(i)} = \begin{bmatrix}
          					\bsK_{t}^{(i)}  & k_t^{(i)}(\bs_{t+1})      \\[0.3em]
      					 k_t^{(i)\top}(\bs_{t+1}) & K^{(i)}(\bs_{t+1}, \bs_{t+1})
					  \end{bmatrix} \]
		\end{itemize}
The three sub-steps above can be performed in parallel across particles.
		\end{itemize}
\indent The correlation range parameters and the latent input could be deterministically propagated by copying them from $S_t^{\zeta(i)}$ to $S_{t+1}^{(i)}$ since they do not change in $t$. Although this strategy is fast, it could lead to particle depletion. To avoid degeneracy in the path space caused by successive resampling steps, we include a ``\textit{rejuvenate}" step which applies MCMC moves to the particles after the propagating step \cite{Gilks2001, Ridgeway2003}. The update is done via elliptical slice sampling \cite{Murray2010}. \\ 
\indent Each particle returns an estimate of predictive mean surface, $\hat{f}^{(i)}$, and predictive standard deviation, $\hat{\sigma}^{(i)}$. Likely, some of these particles will provide higher fidelity surfaces than others. We take the average of the point-wise predictive distribution for each of the particles, the posterior mean predictive curve, as our prediction of $f$ at new inputs 
		\begin{align}
		\hat{f} &= \mathbb{E}({f} \vert S^{(i)}) =  \frac{1}{N}\sum_{i = 1}^N \hat{f}^{(i)}, 
		\end{align}
whereas the estimate for the predictive standard deviation is obtained as
	\begin{equation}
	\sqrt{\hat{\sigma}^2} = \mathbb{E}(\{\hat{\sigma}^{(i)}\}_{i = 1}^N ) + \mathbb{V}\text{ar}(\{\hat{f}^{(i)}\}_{i = 1}^N),
	\end{equation}
\noindent where expressions for $\hat{f}^{(i)}$ and $\hat{\sigma^2}^{(i)}$ are given by Equations (\ref{fhat})--(\ref{sigmahat}), respectively. \\
\indent Sequential MC computing seamlessly blends with active learning of the sequential design (\S \ref{Adaptive sequential design}). In addition to PL, we also investigated a two-stage fast approximation of the proposed emulator where the latent input GP is directly learned from data through nonparametric regression and the estimated input surface is plugged in to learn $f$. Simulations suggest that the two-stage approximation performs at least as well as the sequential MC ``full Bayes" counterpart in handling local features and selecting additional inputs from the boundaries of such features. However, the sequential MC version of our emulator often achieves better accuracy given the same number of input points in the design. A comparison between the two implementations is deferred to Web Appendix D of the Supplementary Materials. 

\subsection{Adaptive sequential design}
\label{Adaptive sequential design}
The adaptive and sequential selection of input points to include in the design sits comfortably within our PL implementation. After particles have been resampled, and before proceeding with the propagation step, the algorithm performs prediction at a set of candidate input configurations based on the posterior predictive distribution. For every candidate point, we derive $\hat{f}(\tilde{\bs} \vert \{\bs,f(\bs)\}_{1:t})$, which predicts $f$ at $\tilde{\bs}$, and $\hat{\sigma}^2(\tilde{\bs} \vert \{\bs, f(\bs)\}_{1:t})$, which quantifies the uncertainty at $\tilde{\bs}$. Candidate points can be ordered based on their predictive variance and the point with largest uncertainty in predicted output is chosen as the next input $\bs_{t+1}$. Consequently, particles are propagated with the new pair $[\bs_{t+1}, f(\bs_{t+1})]$, and the sequence is iterated until some pre-specified stopping criterion is met, e.g. the largest predictive variance falls below a certain threshold or a total number, $T$, of points has been included in the design. In order for this procedure to produce an optimal sequence of points, it is necessary to have a trustworthy judgement of uncertainty, that is, we need to have faith in the model-based estimate of $\hat{\sigma}^2(\tilde{\bs} \vert \{\bs, f(\bs)\}_{1:t})$, which can not be either underestimated or overestimated. Simulation experiments (\S \ref{studies}) show that $\hat{\sigma}^2$ can be poorly estimated by a stationary GP when $f$ presents local features, thus requiring extrinsic diagnostics (e.g., examination of standardised residuals) to help towards the selection of future inputs. To avoid any ad-hoc procedures, it is necessary to rely on an adaptive emulator that can represent properly the simulator. In \S \ref{studies}, we compare a stationary GP to our adaptive emulator in assessing uncertainty in presence of local features. \\
\indent Several authors have developed specific criteria for sequentially selecting new input points. For instance, \cite{Jones1998} proposed an expected improvement criterion to estimate the global mi\-ni\-mum of a computer simulator via the maximum likelihood estimator for the emulator parameters. Equivalently popular approaches are the so-called ``active learning'' criteria such as ALM - active learning MacKay \cite{MacKay1992} and ALC - active learning Cohn \cite{Cohn1996}. The procedure we implement and described above corresponds to ALM. \cite{Seo2000} compared ALM and ALC and observed that ALC often performs better than ALM. For example, the ALM criterion embedded into a stationary GP emulator favours the selection of new points along the boundary of the input space in that the predictive variance is largest beyond the points which are already in the design \cite{MacKay1992}. However, the ALC criterion is more intensive to implement, therefore ALM is often preferred in practice. \\
 
\section{Case studies}
\label{studies}
\subsection{Learning local features}
\begin{figure}[t]
  \centering
  \vspace{-2cm}
  \includegraphics[scale = .6, angle = -90]{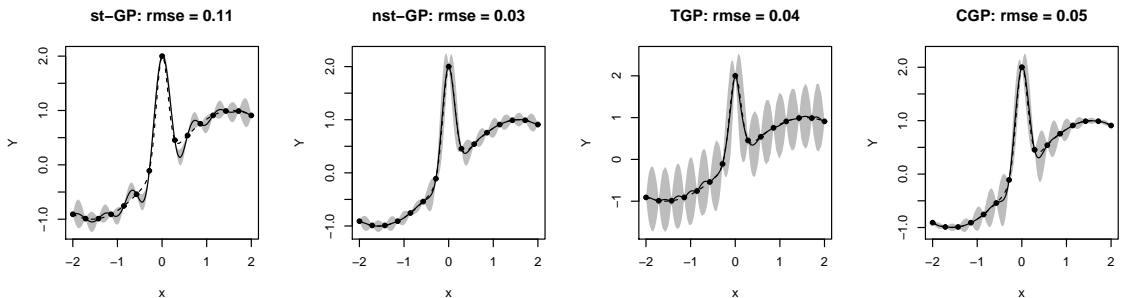}
 \vspace{-2.6cm}
  \caption{Comparison between stationary GP (first panel), non-stationary GP via latent input augmentation (second panel), TGP (third panel), and CGP (fourth panel). The dashed line corresponds to the true function (\ref{peakfun}), the solid black line is the posterior mean predictive curve, and grey areas denote the error bars. Estimates (and RMSE) are obtained at 200 equally spaced test points.} 
   \label{peakfull}
\end{figure}

We consider a spatially inhomogeneous smooth function: 
		\begin{equation}
		f(x) = \sin(x) + 2\exp(-30x^2),
		\label{peakfun}
		\end{equation}
\noindent which is evaluated at 15 equally spaced points in $\Omega = [-2,2]$. \\
\indent For PL, we use $N = 1000$ particles initialised at time $t_0 = 4$ with a randomly selected subset of size 4 of the original 15 points. Each particle contains an estimate of the model parameters, which are initialised by sampling from their prior distributions. $\{\phi_1, \phi_2\}$ and $\tilde{\phi}_1$ are assigned log-normal priors distributions, and 0.5 and 0.25 are chosen as the prior mean and prior variance of the corresponding Normal distribution on $\{\log {\phi}_1, \log {\phi}_2, \log \tilde{\phi}_1\}$. Also, a rather uninformative inverse-gamma prior is chosen for $\sigma^2$, $\sigma^2 \sim \text{IG}(2,1)$. \\
\indent Figure \ref{peakfull} shows the posterior mean predictive curve together with error bars computed as $\hat{f} \pm 2\sqrt{\hat{\sigma}^2}$. We also show results obtained with Bayesian TGP \cite{Gramacy2008} and composite GP (CGP) \cite{Ba2012}. The limitations resulting from fitting a stationary GP to function (\ref{peakfun}) were outlined in Section 1. In comparison, the three non-stationary emulators (panels 2-4 in Figure \ref{peakfull}) give significantly improved performance, i.e. the spline tension effect is eliminated, or strongly attenuated. However, TGP's most evident feature is the large uncertainty in the estimates as quantified by very wide error bars, which could be taken as indicator of an inadequate representation of the simulator. The error bars obtained with our non-stationary GP and CGP are more consistent with the local variability of the underlying surface. In terms of root mean squared error (RMSE), our emulator improves the accuracy of TGP and CPG by 25\% and 40\%, respectively. 

\subsection{Quantifying the emulator's uncertainty} 
The simulator $f$ is ty\-pi\-cal\-ly expected to be within two or three standard deviations from the predictive mean $\hat{f}$ \cite{Bastos2009}. While an isolated outlier might be ignored, several large standardised residuals, e.g. more than 1\% or 5\% of the total number of validating points, may denote a problem to be further investigated. For example, large standardised residuals systematically observed at and around a particular input value suggest that the emulator is not learning the local behaviour of the process \cite{Busby2009}.
Further, they indicate that the emulator is under-estimating the predictive uncertainty. Ultimately, one wants to acquire an accurate knowledge of $f$ with as least simulator's runs as possible. The emulator can be used to quickly identify those regions of the input space where the simulator exhibits more variations, thus help determine where further runs of the simulator should concentrate. However, this goal can be achieved only if the emulator's estimate of uncertainty is trustworthy. A sequential design strategy based on an unreliable estimate of uncertainty will otherwise lead to a sub-optimal selection of input points. \\
 \begin{figure}[t]
  \centering
  \vspace{-.2cm}
  \includegraphics[scale = 0.6, angle = -90]{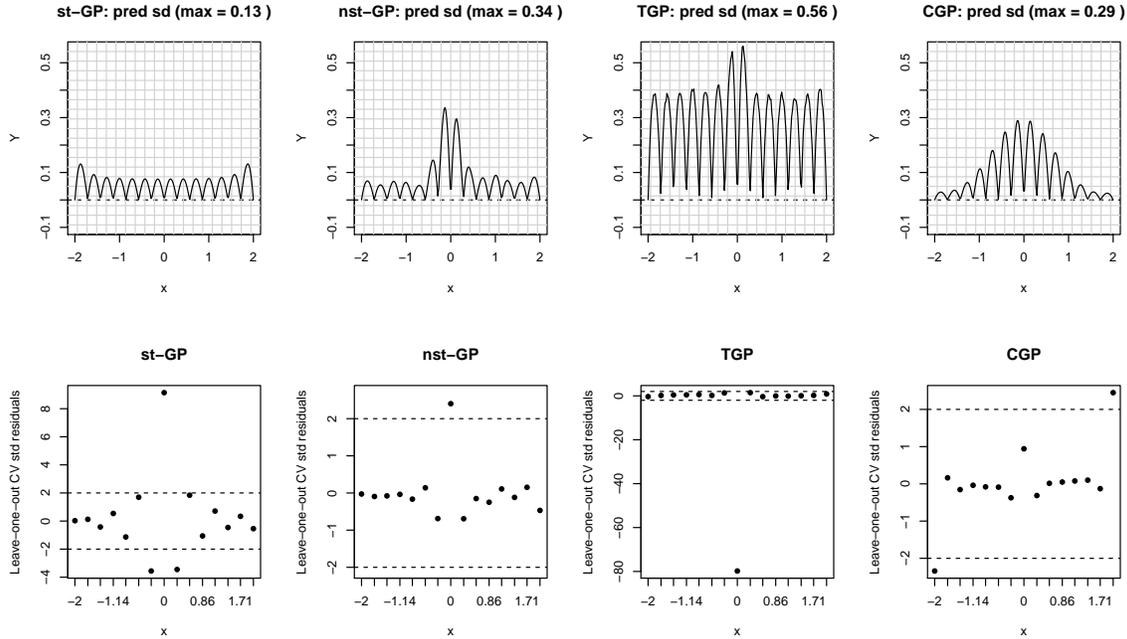}
  \caption{Estimated standard deviation at 200 predictive locations (top panels) and leave-one-out CV standardised residuals (bottom panels) for the peak function (\ref{peakfun}): comparison between stationary GP (st-GP), non-stationary GP via latent input (nst-GP), TGP, and CGP. Top panels also report the maximum estimate of predictive standard deviation.} 
   \label{peakCV}
\end{figure}
\indent Here we examine how model-based evaluations (Figure \ref{peakfull} and first row in Figure \ref{peakCV}) combine with extrinsic diagnostics (second row in Figure \ref{peakCV}). For extrinsic diagnostics, we examine the leave-one-out cross validated standardised residuals. According to the {\textit{exploration}}-driven predictive standard deviation of a stationary GP, one is essentially equally likely to locate the new point anywhere in $[-2, 2]$ (top left panel in Figure \ref{peakCV}). Instead, CV strongly favours the selection of a new input around $x = 0$ ({\textit{exploitation}}-driven CV) to learn the local behaviour of $f$. Thus, model-based evaluations and extrinsic diagnostics are inconsistent, and the latter shows that uncertainty is being under-estimated around the peak. It is not immediately clear how to combine inconsistent evaluations when there is no knowledge of the true $f$, as for real data applications. Incongruent conclusions with CGP: if one trusts the model-based estimate of uncertainty, then the next input will be chosen around the peak; if one relies on CV, the next input will be chosen at the boundaries of the input space. For these two emulators, the problem of how to combine different diagnostic results emerges clearly. Instead, both model-based evaluations and CV for our emulator (nst-GP) and TGP identify that the next point is needed around $x=0$. As opposed to our emulator, the conclusion is however made much more evident by extrinsic diagnostics (a strikingly large CV standardised residual at $x=0$) rather than by the predictive standard deviation with TGP. 

\subsection{1D discontinuous function}
\indent We now consider a simple discontinuous function:
		\begin{equation}
	        f(x) = \left\{
		 \begin{array}{l l}
    			0, & \quad x \leq 0\\
   			1, & \quad x > 0\\
              \end{array} \right.
	       \label{jump}
	       \end{equation}

\noindent We evaluate (\ref{jump}) at 10 equally spaced points in $\Omega = [-1,1]$. For PL, we use $N = 1000$ particles initialised at time $t_0 = 4$ with a randomly selected subset of size 4 of the original 10 points. This function is particularly suited to TGP because of the vertical, axis-aligned nature of localised feature. \\
 \begin{figure}[t]
  \centering
  \vspace{-2cm}
  \includegraphics[scale = .60, angle = -90]{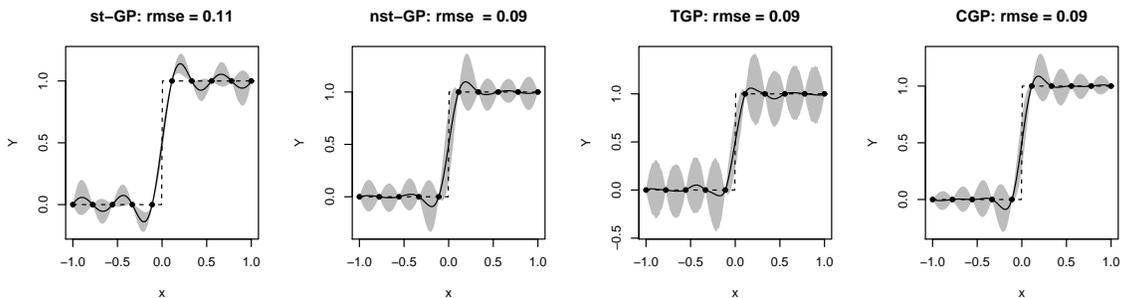}
  \vspace{-2.5cm}
  \caption{Comparison between stationary GP (st-GP), non-stationary GP via latent input augmentation (nst-GP), TGP, and CGP on the 1D discontinuous function. The dashed line corresponds to the true function (\ref{jump}), the solid black line is the posterior mean predictive curve, and grey areas denote the error bars. Estimates (and RMSE) are obtained at 200 equally spaced test points.}
   \label{jumpfigure}
   \end{figure}
\indent Figure \ref{jumpfigure} shows that the point predictions obtained with the non-stationary emulators are not (or less) distorted by the spline tension effect as opposed to the fit obtained with a stationary GP. Further, the intervals seem more consistent with what might be guessed about the function from observing the data points. Again, TGP identifies large uncertainty everywhere in $\Omega$. \\
 \begin{figure}[h]
  \centering
  \includegraphics[scale = .6, angle = -90]{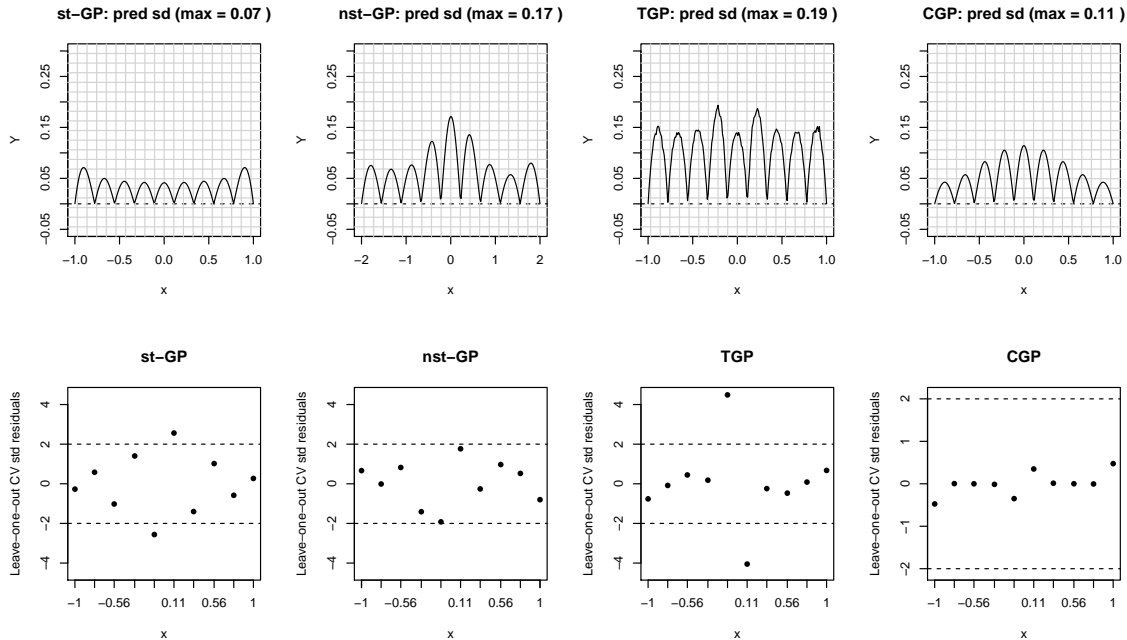}
  \caption{Estimated standard deviation at 200 predictive locations (top panels) and leave-one-out CV standardised residuals (bottom panels) for the jump function (\ref{jump}): comparison between stationary GP (st-GP), non-stationary GP via latent input (nst-GP), TGP, and CGP. Top panels also report the maximum estimate of predictive standard deviation.} 
   \label{jumpfigure2}
   \end{figure}
\indent Model-base evaluations for our non-stationary GP and CGP suggest to pick new points at (and around) $x = 0$ to exploit the local feature (top row in Figure \ref{jumpfigure2}). Also, CV shows that the predictive uncertainty is not underestimated. Extrinsic and model-based evaluations are still inconsistent for the stationary GP, whereas extrinsic diagnostics show that TGP is likely to be under-estimating the uncertainty at the jump.  Although this would lead to a sequential design strategy consistent with the one suggested by TGP's model-based evaluation, it is preferable to observe the more apparent pattern in our emulators's predictive standard deviation, which drops quickly when departing from $x=0$. \\
\indent In general, it is not clear how to reconcile model-based and extrinsic diagnostics whenever these lead to different evaluations. In particular, it is not obvious in what measure to favor the {\textit{exploration}-}driven predictive standard deviation over the {\textit{exploitation}-}driven CV.  An emulator whose model-based evaluations reconcile with extrinsic diagnostics is preferred in that it automatically learns to create a good balance between {\textit{exploration}} and {\textit{exploitation}}, and one does not have to resort to ad-hoc combinations. Our emulator seems to accomplish this balance adequately. 

\section{High-dimensional examples and sequential design}
\label{sequential higher}
\subsection{Two-dimensional functions with local features}
\label{two-D}
\begin{figure}[t]
  \centering
  \vspace{-2cm}
  \includegraphics[scale=.70, angle = -90]{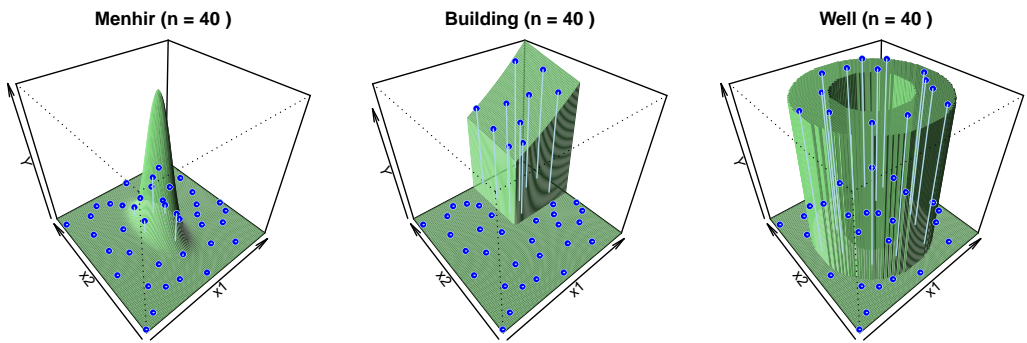}
    \vspace{-3cm}
  \caption{True functions for the 2-dimensional numerical examples and initial 40 LHD (blue points) used to train the emulators.}
   \label{true}
\end{figure}

\begin{figure}[h!]
  \centering
  \includegraphics[scale = .53, angle = -90]{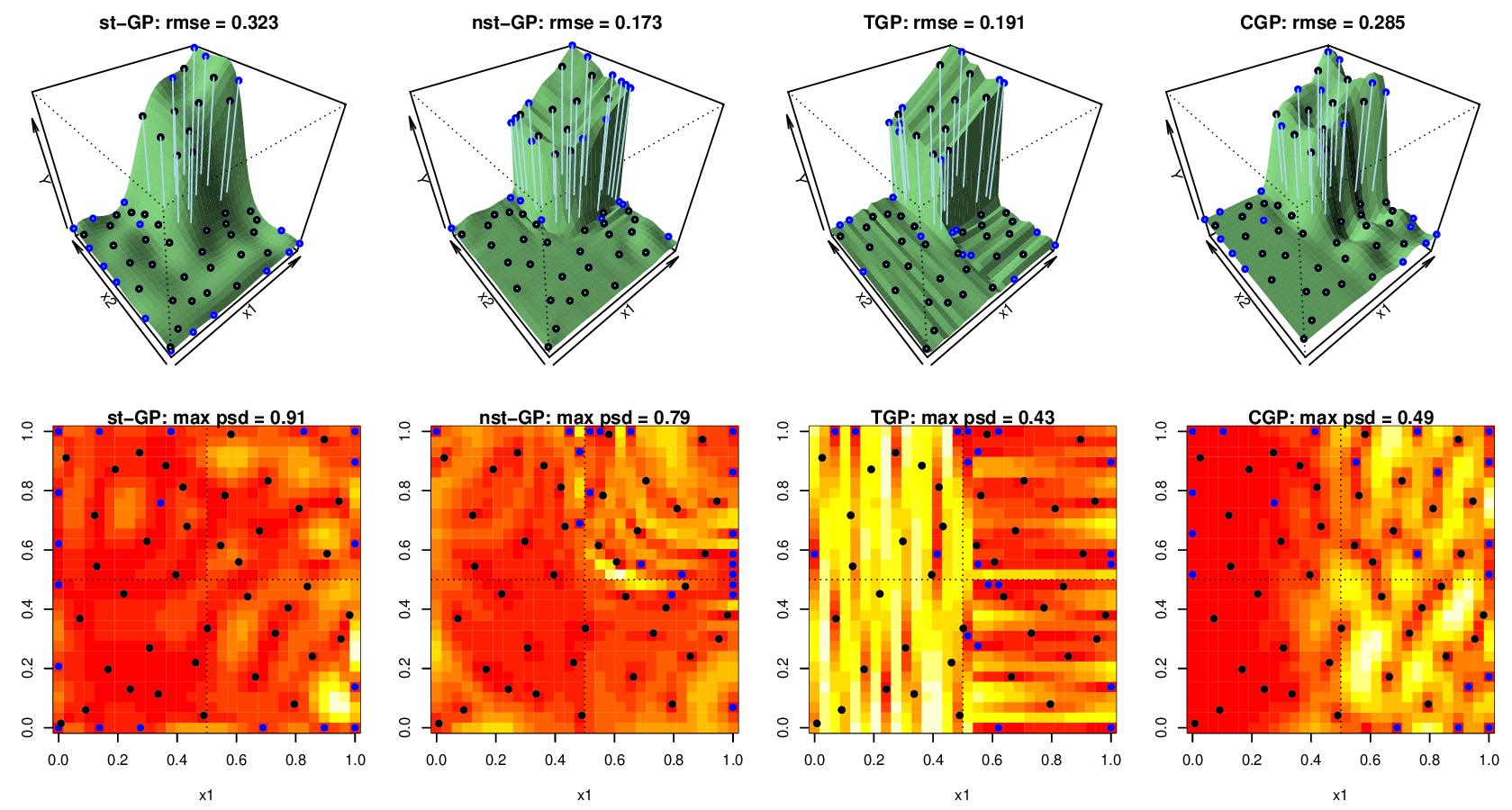}
  \caption{Comparison between stationary GP (st-GP), non-stationary GP via latent input augmentation (nst-GP), TGP, and CGP on the 2D building function at $T = 60$. Black points: initial 40 LHD (common to all emulators); blue points: 20 additional points selected via ALM. Top row: posterior mean predictive surface, $\hat{f}$, and root mean squared error (rmse); bottom row: predictive standard deviation, $\hat{\sigma}$, and maximum predictive standard deviation (max psd). The quality of the prediction is assessed at a collection of 900 points in $\Omega = [0,1]^2$, i.e. an expanded grid of 30 equally spaced points along each coordinate axes.}
   \label{nstat5}
\end{figure}

In this Section, we examine three test functions possessing non-stationary features. The true surfaces are shown in Figure \ref{true}: the second function (``buil\-ding'') is naturally suited to TGP because of the axis-aligned non-stationarity. \\
\indent We first compare the performance of the emulators when trained on a common and fixed set of input points (no sequential design). For this purpose, we use a 40 LHD (blue points in Figure \ref{true}), which allows the emulators to gather knowledge on the overall shape of $f$ because of its ``space-filling" nature. Figure \ref{menhir40} in Appendix \ref{Menhir40} and Figures \ref{nstat2}-\ref{nstat3} show the posterior predictive mean surface, $\hat{f}$, and the predictive standard deviation, $\hat{\sigma}$, for the menhir, building, and well functions, respectively. The initial LHD does not include points at or nearby the peak of the menhir function (Figure \ref{menhir40}), and this affects the estimates of the four emulators, which can not recover the central spike. Note, however, how both our emulator and CGP identify higher uncertainty in the central part of the input space. In particular, the pattern in predictive standard deviation seems to indicate that our emulator is learning the circular geometry of the menhir. TGP also identifies higher central uncertainty, but the reason is likely to be related to the partitioning scheme rather than to learning the geometry of the feature. The most distinctive feature that emerges from both Figure \ref{nstat2} and Figure \ref{nstat3} is that our emulator is learning the geometry of the local features as shown by the evident patterns in predictive standard deviation, which is higher at the edges of the building (Figure \ref{nstat2}) and has a distinctive circular pattern for the well function (\ref{nstat3}). This does not appear to be the case for the other emulators. \\
  \begin{figure}[t]
  \centering
  \includegraphics[scale = .53, angle = -90]{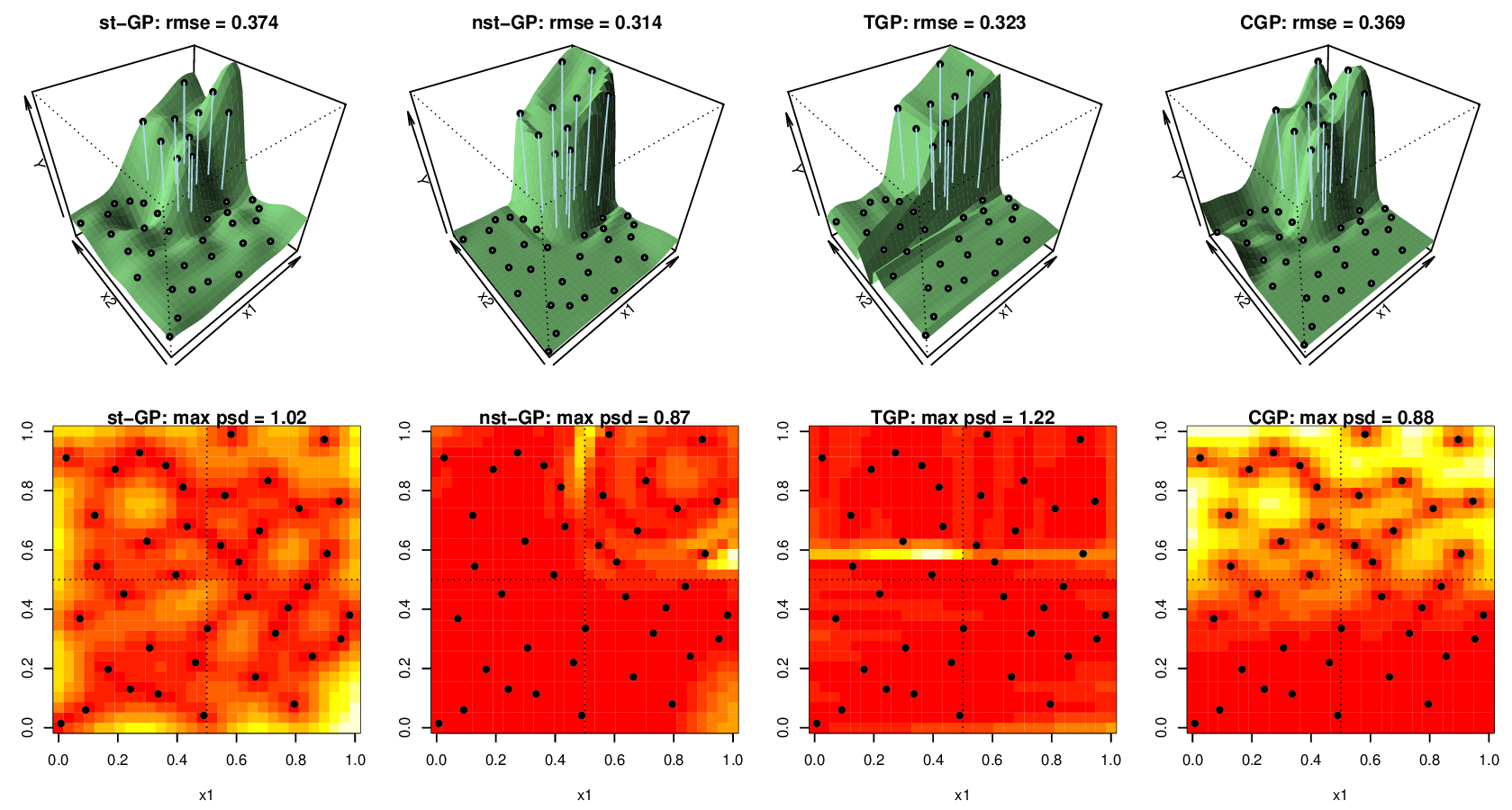}
  \caption{Comparison between stationary GP (st-GP), non-stationary GP via latent input augmentation (nst-GP), TGP, and CGP on the 2D building function. The emulators are fit to a common design corresponding to a 40 LHD (black points). Top row: posterior mean predictive surface, $\hat{f}$, and root mean squared error (rmse); bottom row: predictive standard deviation, $\hat{\sigma}$, and maximum predictive standard deviation (max psd). The quality of the prediction is assessed at a collection of 900 points in $\Omega = [0,1]^2$, i.e. an expanded grid of 30 equally spaced points along each coordinate axes.}   
  \label{nstat2}
\end{figure}
  \begin{figure}[h!]
  \centering
  \includegraphics[scale = .53, angle = -90]{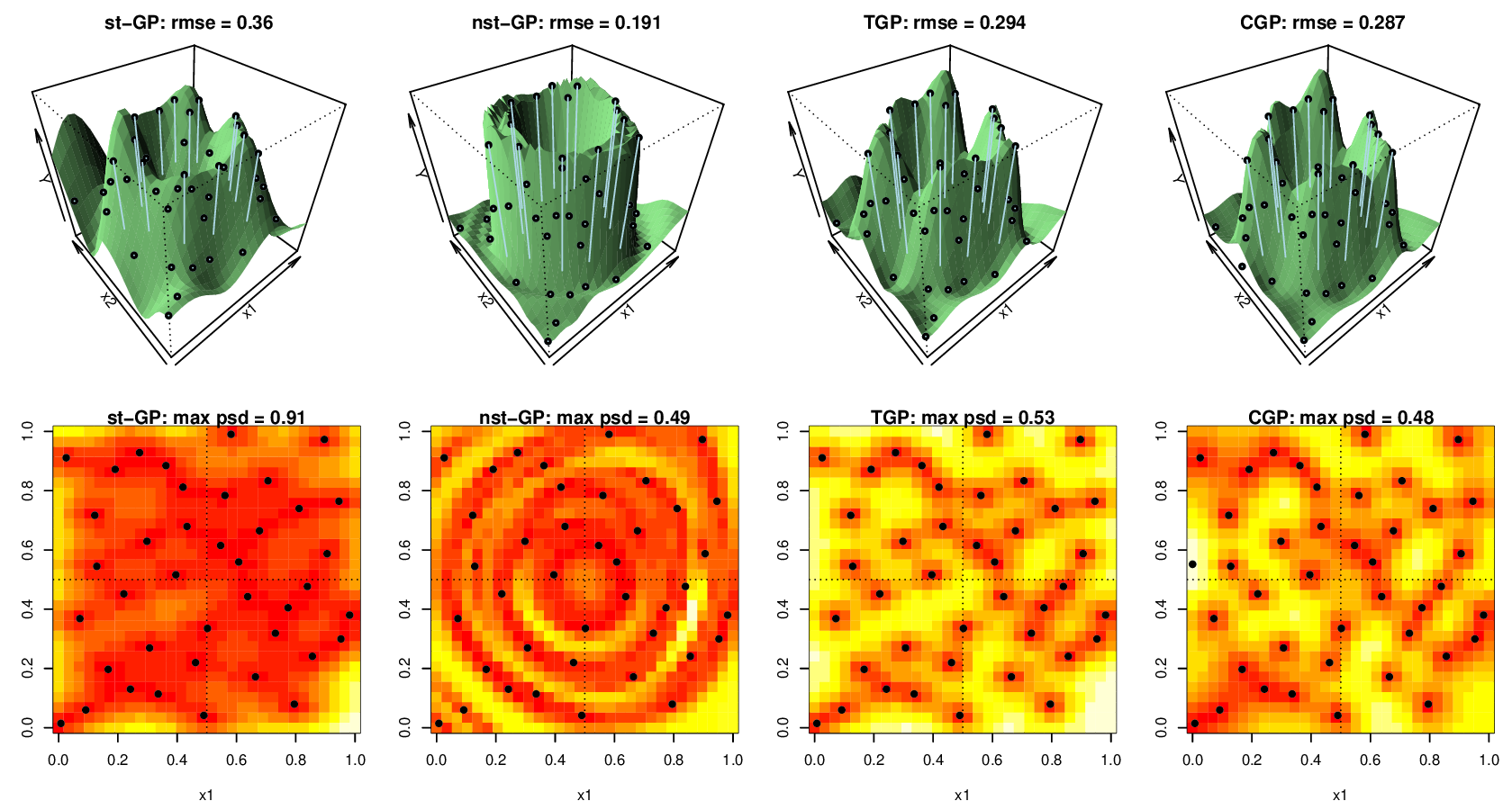}
  \caption{Comparison between stationary GP (st-GP), non-stationary GP via latent input augmentation (nst-GP), TGP, and CGP on the 2D well function. The emulators are fit to a common design corresponding to a 40 LHD (black points). Top row: posterior mean predictive surface, $\hat{f}$, and root mean squared error (rmse); bottom row: predictive standard deviation, $\hat{\sigma}$, and maximum predictive standard deviation (max psd). The quality of the prediction is assessed at a collection of 900 points in $\Omega = [0,1]^2$, i.e. an expanded grid of 30 equally spaced points along each coordinate axes.}
   \label{nstat3}
\end{figure}
\indent Next, we address the issue of sequential design and assess whether the emulators can correct for inadequacies in the fit. In other terms, we want to examine whether the emulators can learn about, and thus concentrate exploration in, the most interesting or complicated regions of the input space. Therefore, we let the emulators select 20 additional points (60 for well) sequentially based on their model-based estimate of uncertainty (ALM). The resulting final designs will, therefore, be different across emulators. \\
\indent Figures S1 and S2 in Web Appendix A of the Supplementary Materials show $\hat{f}$ and $\hat{\sigma}$ for the menhir and well functions at $T = 60$ and $T = 100$, respectively, and Figure \ref{nstat5} refers to the building function at $T = 60$. Regardless of the function being examined, our emulator favours the sampling of new points from the boundaries of the features. Therefore, it strikes a good balance between {\textit{exploration}} (initial LHD) and {\textit{exploitation}} (newly selected points). This is not necessarily true for the other emulators across different functions, i.e. CGP tends to select new points at the center of the input space for the menhir function, but no pattern is observed for building and well functions. After ALM, TGP seems to have preferred the selection of new points at the edges of the building (Figure \ref{nstat5}), but this behaviour is much less evident on the other functions with circular geometry. TGP's selection is driven by the partitioning scheme in that the predictive standard deviation is generally higher at the edges between consecutive partitions. Thus, TGP seems to concentrate in learning the partition rather than the local feature. \\ 
 \begin{figure}[h!]
\vspace{-2cm}
  \centering
  \includegraphics[scale = .61, angle = -90]{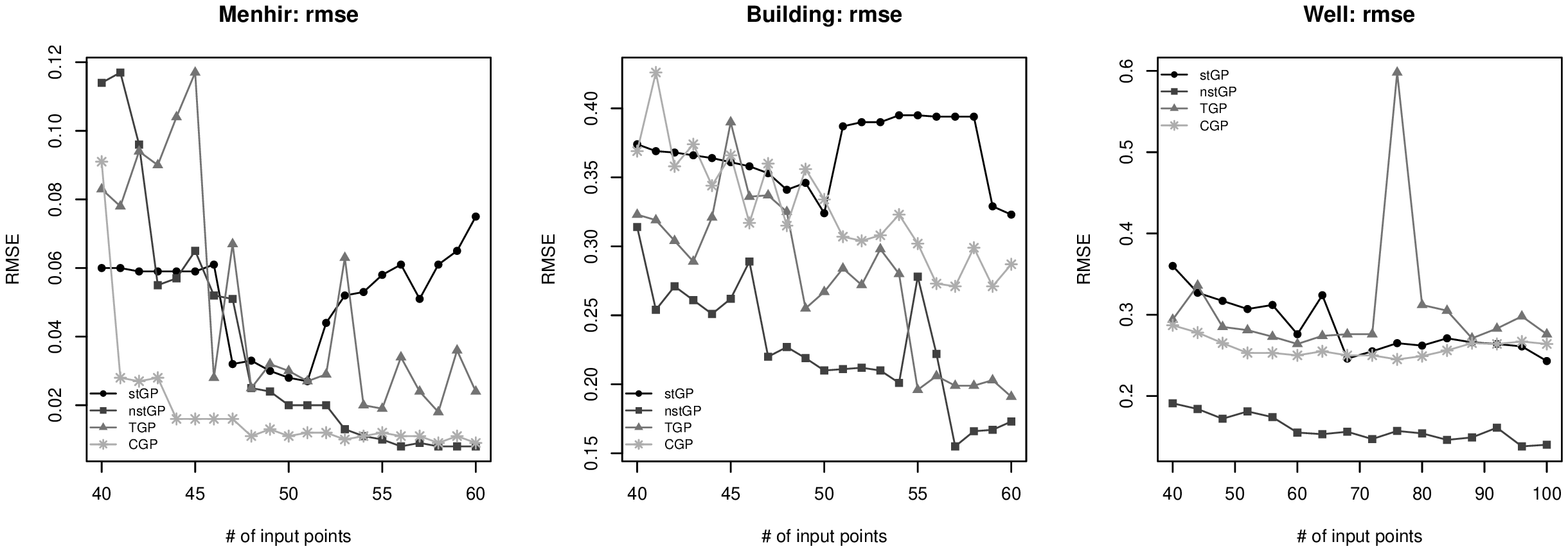}
  \vspace{-2.2cm}
    \caption{Progression of the RMSE as additional input points are being selected for the 2D functions.}
   \label{treed4}
\end{figure}
\indent For a more quantitative numerical comparison among emulators, Figure \ref{treed4} shows the progression of the RMSE as additional inputs are being selected. Our emulator performs at least as well as CGP on the menhir function, and outperforms the other emulators on building and, in particular, well functions. \\ 
\indent To conclude, our emulator tends to concentrate the selection of new points in interesting areas of the input space. Furthermore, it compares favourably both in cases of axis-aligned non-stationarity (building) and in situations where the type of non-stationarity is more general (well). 

\subsection{Six-dimensional examples}
\begin{figure}[t]
  \vspace{-1.6cm}
  \centering
  \includegraphics[scale = .6, angle = -90]{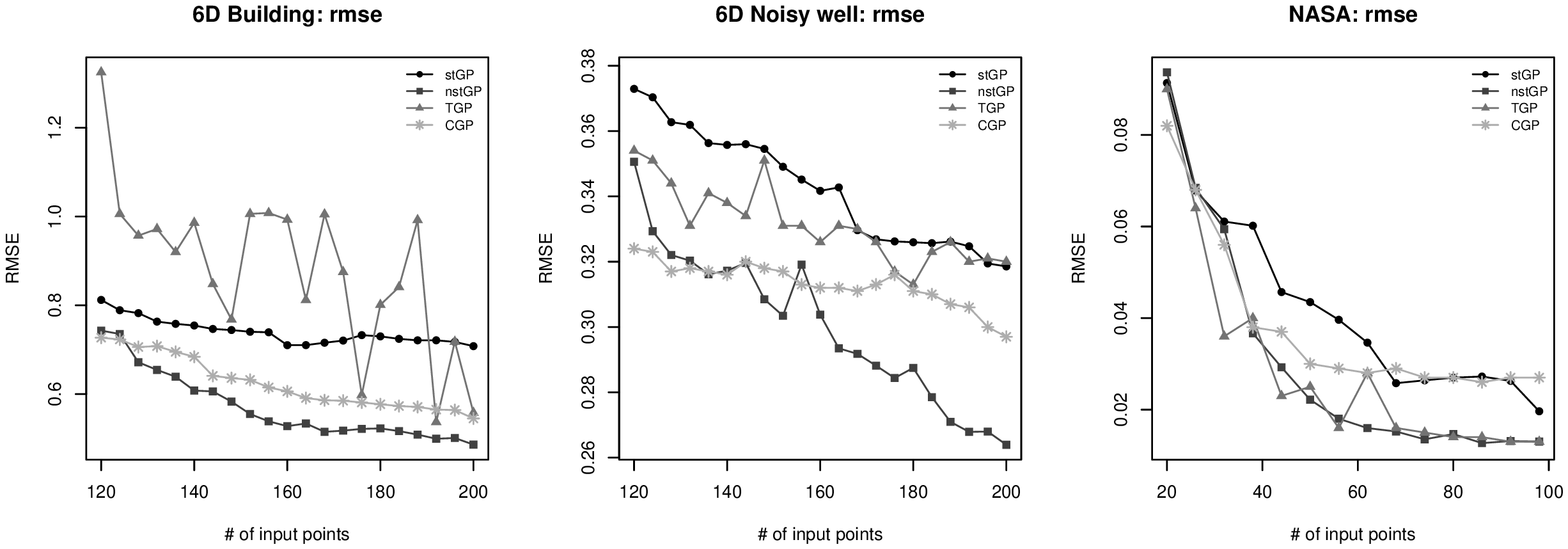}
   \vspace{-2cm}
  \caption{Comparison among stationary GP (stGP), non-stationary GP via latent input augmentation (nstGP), TGP, and CGP in terms of progression of the RMSE as additional input points are being selected. Left panel: 6D building function (\ref{nbuilding}). Central panel: 6D well (\ref{nwell}); Right panel: LGBB CFD experiment (Section \ref{LGBB}).} 
     \label{6drmse}
\end{figure}

We consider two 6D examples, which are an extension of the 2D building and well functions. The 6D building has true function:
	\begin{equation}
	f(x_1, x_2, x_3, x_4, x_5, x_6) = 
		\left\{
		 \begin{array}{l l}
    			e^{\sum_{i = 1}^ 6 \left(\frac{1}{i}\right)^2 x_i}, & \quad \text{if } x_1, x_2, x_3, x_4, x_5, x_6 > 0.25\\
   			0, & \quad \text{otherwise}
			\label{nbuilding}
              \end{array} \right.	\end{equation}
on the hypercube $\boldsymbol{X} = [0,1]^6$. The 6D well has true function:
	\begin{equation}
	f(x_1, \dots, x_6) = 
		\left\{
		 \begin{array}{l l}
    			1, & \quad \text{if }\sum_{i = 1}^4 (x_i - 0.5)^2  > 0.025 \text{ and }  \sum_{i = 1}^4 (x_i - 0.5)^2  < 0.25\\
   			0, & \quad \text{otherwise}
			\label{nwell}
              \end{array} \right.	\end{equation}
on the hypercube $\boldsymbol{X} = [0,1]^6$. Therefore, $f$ in (\ref{nwell}) is constant in $x_5$ and $x_6$. \\ 
\indent All emulators are trained on an identical 120 LHD. Emulators then select 80 additional points from a 1000 candidate LHD according to ALM. Similar to the 2D examples, our emulator outperforms the others in terms of reduction of RMSE (left and central panels in Figure \ref{6drmse}). Therefore, inactive covariates adding noise to the process, such as $x_5$ and $x_6$ for function (\ref{nwell}), do not affect the performance of our emulator. Additional summaries are reported in Web Appendix B of the Supplementary Materials.\\ 

\section{LGBB CFD experiment}
\label{LGBB}
\begin{figure}[h]
  \centering
 \includegraphics[scale = .55, angle = -90]{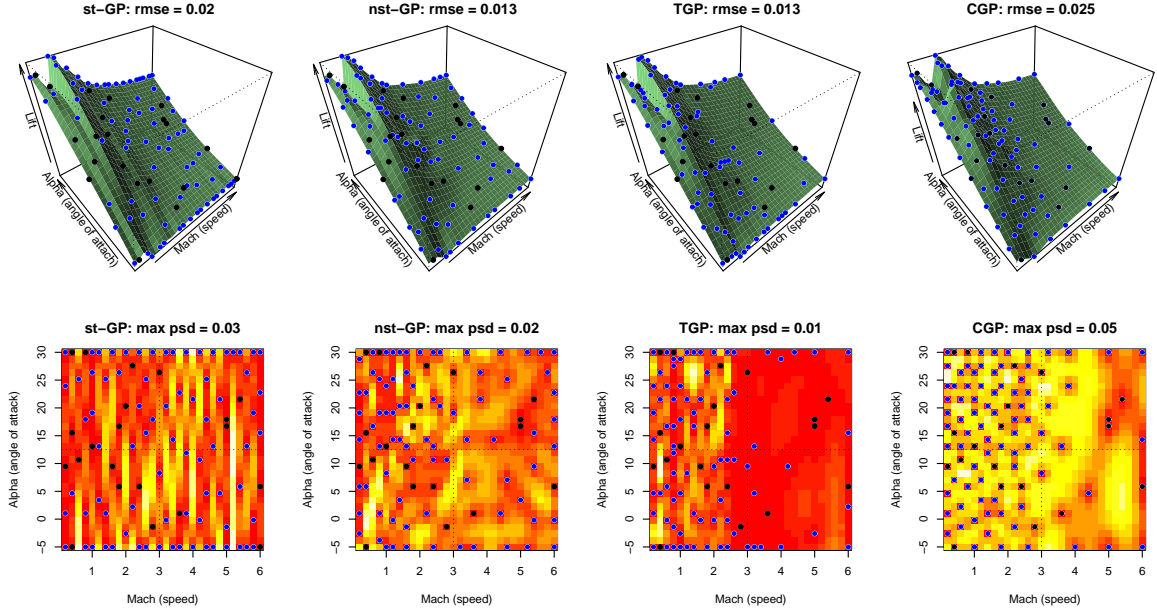}
  \caption{Comparison between stationary GP (st-GP), non-stationary GP via latent input augmentation (nst-GP), TGP, and CGP on the LGBB experiment with lift response as a function of Mach (speed) and Alpha (angle of attack) and Beta (side-slip angle) fixed at 0. Black points: 20 points used as initial design (common to all emulators); blue points: 80 additional points selected via ALM. Top row: posterior mean predictive surface, $\hat{f}$, and root mean squared error (rmse); bottom row: predictive standard deviation, $\hat{\sigma}$, and maximum predictive standard deviation (max psd). The quality of the prediction is assessed at a collection of 900 points, i.e. an expanded grid of 30 equally spaced points along each coordinate axes.}
   \label{nasa}
\end{figure}

This Section presents an application to a computational fluid dynamics (CFD) simulator of a proposed reusable NASA rocket booster vehicle, the Langley Glide-Back Booster (LGBB). The interest is in learning about the response in several flight characteristics of the LGBB as a function of three inputs (speed in Mach number, angle of attack, and slide-slip angle) when the vehicle reenters the atmosphere. See \cite{Gramacy2009} for more details on the study. \\ 
\indent The CFD simulation involves the iterative integration of systems of inviscid Euler equations and each run of the solver for a given set of parameters takes on the order of 5--20 hours on a high-end workstation \cite{Gramacy2009}. Therefore, the interest in adaptively design the experiment to concentrate sampling in those regions where the response is more interesting (e.g., higher uncertainty or richest structure) emerges clearly. As \cite{Gramacy2009} show, the most interesting region occurs near Mach 1 and for large angle of attack (refer to Figure S5 in Web Appendix C of the Supplementary Materials for a plot of the ``lift" response as function of Mach and Alpha). The ridge in response at Mach equal to 1 separates subsonic flows and supersonic flows. The behaviour of the response is quite different in the two regions, with lift appearing mostly homogeneous in the supersonic region. \\ 
\indent Following \cite{Gramacy2009}, we examine the lift response as a function of speed (Mach) and angle of attach (Alpha) with the side-slip angle (Beta) fixed at zero. We obtain a linear interpolation onto a $30 \times 30$ grid over Mach and Alpha, and use the interpolated lift as our truth. All emulators are trained on a fixed and common initial design given by 20 randomly selected points from a $30 \times 30$ grid (Mach $\in [0, 6]$, Alpha $\in [-5, 30]$), then select 80 additional points via ALM. Figure \ref{nasa} shows a slice of the posterior mean predictive surface as a function of Mach and Alpha. The distinction between subsonic and supersonic flows is well captured by all non-stationary emulators, which tend to select new points at small Mach, and our emulator and CGP do so particularly for large Alpha. A stationary GP focuses mostly on a uniform exploration of the space and will require ad-hoc extrinsic diagnostics to focus around the ridge. \\ 
\indent The third panel in Figure \ref{6drmse} shows the progression of RMSE to the interpolated truth. Our emulator performs as well as TGP on a surface that favours the latter because of the axis-aligned local feature, and improves the accuracy over stationary GP and CGP by 35\% and 48\%, respectively, at $T = 100$.

\section{Discussion}
\label{conclusions}
In this work we describe a non-stationary GP model that can be used as an emulator in the sequential design of computer experiments. To induce non-stationarity, we consider a mapping to a latent space where stationarity holds, and augment the input space by the latent input. The numerical examples show that the extra flexibility introduced by the latent input greatly improves predictions over a stationary GP fit. In particular, the proposed methodology provides more reliable, model-based evaluations as opposed to extraneous explorations done with stationary GPs, and adapts to both cases of axis-aligned non-stationarity and in situations where the non-stationarity is more general. The approach also retains an easy interpretability while building upon a simple but elegant construction. Here we discuss some details in regard to our implementation and computer emulation in general. \\
\indent {\it{The nugget}.} A ``nugget" is a small, positive quantity $\alpha$ often added to the diagonal of the correlation function for $f$ \cite{Andrianakis2012}. The resulting covariance function corresponds to the case where $f$ is observed with additive Gaussian noise with zero mean and variance $\alpha$. Many authors do not include a nugget term on the grounds that computer codes are deterministic. In fact, the nugget introduces a measurement error in the stochastic process. A GP that includes a nugget does not interpolate and assigns non-zero uncertainty to the design data. However, it is not uncommon practice to include a nugget to enhance the numerical stability in factorising covariance matrices \cite{Gramacy2008, Andrianakis2012}. A typical value of the nugget used in our numerical examples is $\alpha = 10^{-7}$, the effect of this being the addition of $\alpha$ in the predictive variance of the responses. Although very small, $\alpha$ can have a non-negligible impact on the estimates. For example, it compromises interpolation of the stationary GP on the menhir function (Figure S1 in Web Appendix A). However, the nugget did not seem to significantly affect the estimates of our non-stationary emulator. For more details on the inclusion of a nugget in computer emulation, refer to \cite{Andrianakis2012, Ba2012}. \\
\indent {\it{Parallelisation}.} An appealing characteristic of PL is its heavy parallelisable nature: many of the typical calculations on the particles can proceed independently of one another. In particular, the evaluation of the posterior predictive distribution and the propagate step can be performed in parallel for each particle. Resampling requires the particles being synchronised, but this step is fast once the particle predictive densities have been evaluated. Our code makes extensive use of {\url{R}}'s function {\url{lapply}} to automatically loop over the particles to evaluate the predictive distribution and propagate the particles. More advanced alternatives such as {\url{snowfall}} and {\url{sfCluster}} for parallel programming using clusters could lead to computational improvements. \\
\indent {\it{High-dimensional problems}.} The application of the proposed methodology to high-dimensional input spaces can be challenging due to the intrinsic difficulty faced in high-dimensional settings by GP models, which try to recover up to the $p$-$th$ level of interaction. We expect that more structure (i.e., additivity, sparse factorisation) is needed to handle high-dimensional problems. Independently developed research in the context on non-parametric regression suggests that additive GP models could be a promising way to move forward, and they will be investigated in future research.  \\
\indent {\it{Applications}.} Although the model was developed for the analysis of computer experiments, it also has a wide range of uses as a simple and efficient method for non-stationary modelling in the analysis of social, biological, and ecological data collected over spatial domains. The extension to non-parametric regression is straightforward with the inclusion of a nugget \cite{Schmidt2000, Rasmussen2001, Kim2005, Paciorek2006}.  

\appendix
\section{Menhir function estimates at $T = 40$ (LHD)} 
\label{Menhir40}
 \begin{figure}[htb!]
  \centering
 \includegraphics[scale=.53, angle = -90]{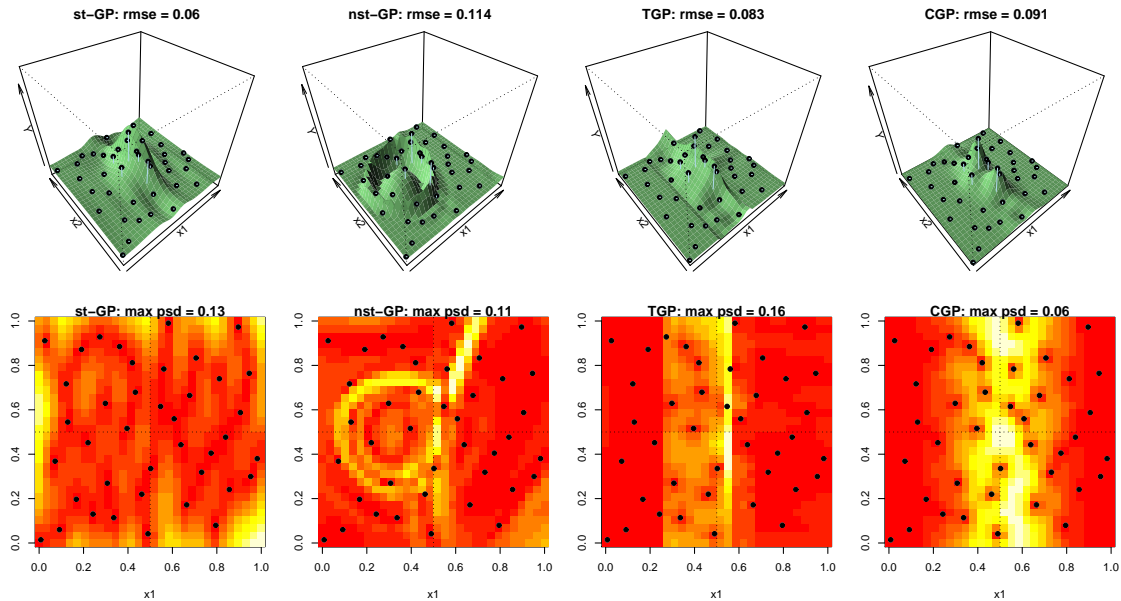}
\caption{Comparison between stationary GP (st-GP), non-stationary GP via latent input augmentation (nst-GP), TGP, and CGP on the 2D menhir function. The emulators are fit to a common design corresponding to a 40 LHD (black points). Top row: posterior mean predictive surface, $\hat{f}$, and root mean squared error (rmse); bottom row: predictive standard deviation, $\hat{\sigma}$, and maximum predictive standard deviation (max psd). The quality of the prediction is assessed at a collection of 900 points in $\Omega = [0,1]^2$, i.e. an expanded grid of 30 equally spaced points along each coordinate axes.}
\label{menhir40}
\end{figure}


\bibliographystyle{siam}
 \bibliography{samsi.bib}

\begin{thebibliography}{10}

\bibitem{Andrianakis2012}
{\sc I.~Andrianakis and P.~G. Challenor}, {\em {The effect of the nugget on
  Gaussian process emulators of computer models}}, Comput. Stat. Data Anal., 56
  (2012), pp.~4215--4228.

\bibitem{Ba2012}
{\sc S.~Ba and R.~Joseph}, {\em {Composite Gaussian process models for
  emulating expensive functions}}, Annals of Applied Statistics, 6 (2012),
  pp.~1838--1860.

\bibitem{Bastos2009}
{\sc L.~S. Bastos and A.~O'Hagan}, {\em {Diagnostics for Gaussian process
  emulators}}, Technometrics,, 51 (2009), pp.~425--438.

\bibitem{Bayarri2007}
{\sc M.~J. Bayarri, J.~O. Berger, R.~Paulo, J.~Sacks, J.~A. Cafeo,
  J.~Cavendish, C.-H. Lin, and J.~Tu}, {\em A framework for validation of
  computer models}, Technometrics, 49 (2007), pp.~138--154.

\bibitem{Bhattacharya2011}
{\sc A.~Bhattacharya, D.~Pati, and D.~B. Dunson}, {\em {Adaptive dimension
  reduction with a Gaussian process prior}}, arXiv: 1111.1044,  (2011).

\bibitem{Busby2009}
{\sc D.~Busby}, {\em {Hierarchical adaptive experimental design for Gaussian
  process emulators}}, Reliability Engineering \& System Safety, 94 (2009),
  pp.~1183--1193.

\bibitem{Cohn1996}
{\sc D.~A. Cohn}, {\em Neural network exploration using optimal experiment
  design}, Neural Netw.,, 9 (1996), pp.~1071--1083.

\bibitem{Currin1991}
{\sc C.~Currin, T.~Mitchell, M.~Morris, and D.~Ylvisaker}, {\em {Bayesian
  prediction of deterministic functions, with applications to the design and
  analysis of computer experiments}}, J. Am. Statist. Assoc., 86 (1991),
  pp.~953--963.

\bibitem{Fan2009}
{\sc Y.~Fan, I.~Ginis, T.~Hara, C.~W. Wright, and E.~J. Walsh}, {\em {Numerical
  simulations and observations of surface wave fields under an extreme tropical
  cyclone}}, J. Phys. Oceanogr., 39 (2009), pp.~2097--2116.

\bibitem{Gilks2001}
{\sc W.~R. Gilks and C.~Berzuini}, {\em {Following a moving target: Monte Carlo
  inference for dynamic Bayesian models}}, J. R. Statist. Soc. B, 63 (2001),
  pp.~127--146.

\bibitem{Gramacy2008}
{\sc R.~B. Gramacy and H.~K.~H. Lee}, {\em {Bayesian treed Gaussian process
  models with an application to computer modeling}}, J. Am. Statist. Assoc.,
  103 (2008), pp.~1119--1130.

\bibitem{Gramacy2009}
{\sc R.~B. Gramacy and H.~K.~H. Lee}, {\em {Adaptive design and analysis of
  supercomputer experiments}}, Technometrics, 51 (2009), pp.~130--145.

\bibitem{Gramacy2011}
{\sc R.~B. Gramacy and N.~G. Polson}, {\em {Particle learning of Gaussian
  process models for sequential design and optimization}}, J. Comput. and
  Graph. Statist., 20 (2011), pp.~102--118.

\bibitem{Jones1998}
{\sc D.~R. Jones, M.~Schonlau, and W.~J. Welch}, {\em Efficient global
  optimization of expensive black-box functions}, J. of Global Optimization, 13
  (1998), pp.~455--492.

\bibitem{Kennedy2001}
{\sc M.~C. Kennedy and A.~O'Hagan}, {\em Bayesian calibration of computer
  models}, J. R. Stat. Soc. B, 63 (2001), pp.~425--464.

\bibitem{Kim2005}
{\sc H.-M. Kim, B.~K. Mallick, and C.~Holmes}, {\em {Analyzing nonstationary
  spatial data using piecewise Gaussian processes}}, J. A. Stat. Assoc., 100
  (2005), pp.~653--668.

\bibitem{Lopes2011}
{\sc H.~F. Lopes, C.~M. Carvalho, M.~S. Johannes, and N.~G. Polson}, {\em
  Particle learning for sequential Bayesian computation}, Oxford University
  Press, 2011.

\bibitem{MacKay1992}
{\sc D.~J. MacKay}, {\em Information-based objective functions for active data
  selection}, Neural Computation, 4 (1992), pp.~590--604.

\bibitem{Murray2010}
{\sc I.~Murray, R.~P. Adams, and D.~J.~C. MacKay}, {\em {Elliptical slice
  sampling}}, J. Mach. Learn. Res., 9 (2010), pp.~541--548.

\bibitem{Paciorek2004}
{\sc C.~J. Paciorek and M.~J. Schervish}, {\em Nonstationary covariance
  functions for gaussian process regression}, in Proc. of the Conf. on Neural
  Information Processing Systems (NIPS), vol.~16, MIT Press, 2004,
  pp.~273--280.

\bibitem{Paciorek2006}
\leavevmode\vrule height 2pt depth -1.6pt width 23pt, {\em Spatial modelling
  using a new class of nonstationary covariance functions}, Environmetrics, 17
  (2006), pp.~483--506.

\bibitem{Pfingste2006}
{\sc T.~Pfingsten, M.~Kuss, and C.~E. Rasmussen}, {\em {Nonstationary Gaussian
  process regression using a latent extension of the input space}}, in
  manuscript, 2006.

\bibitem{Rasmussen2001}
{\sc C.~E. Rasmussen and Z.~Ghahramani}, {\em {Infinite mixtures of Gaussian
  process experts}}, in Advances in Neural Information Processing Systems,
  vol.~14, MIT Press, 2001, pp.~881--888.

\bibitem{Ridgeway2003}
{\sc G.~Ridgeway and D.~Madigan}, {\em {A sequential Monte Carlo method for
  Bayesian analysis of massive datasets}}, J. Knowl. Disco. and Data Min., 7
  (2003), pp.~301--319.

\bibitem{Rougier2009}
{\sc J.~Rougier, S.~Guillas, A.~Maute, and A.~D. Richmond}, {\em Expert
  knowledge and multivariate emulation: the thermosphere-ionosphere
  electrodynamics general circulation model (tie-gcm)}, Technometrics, 51
  (2009), pp.~414--424.

\bibitem{Sacks1989}
{\sc J.~Sacks, W.~J. Welch, T.~J. Mitchell, and H.~P. Wynn}, {\em {Design and
  analysis of computer experiments}}, Statist. Sci., 4 (1989), pp.~409--423.

\bibitem{Sampson1992}
{\sc P.~D. Sampson and P.~Guttorp}, {\em {Nonparametric estimation of
  nonstationary spatial covariance structure}}, J. A. Stat. Assoc., 87 (1992),
  pp.~108--119.

\bibitem{Santner2003}
{\sc T.~Santner, B.~Williams, and W.~Notz}, {\em The Design and Analysis of
  Computer Experiments}, Springer Series in Statistics, Springer, 2003.

\bibitem{Schade1999}
{\sc L.~Schade and K.~Emanuel}, {\em {The ocean's effect on the intensity of
  tropical cyclones: results from a simple coupled atmosphere-ocean model}}, J.
  Atmos. Sci., 56 (1999), pp.~642--651.

\bibitem{Schmidt2000}
{\sc A.~M. Schmidt and A.~O'Hagan}, {\em Bayesian inference for nonstationary
  spatial covariance structure via spatial deformations}, J. R. Stat. Soc., B,
  65 (2000), pp.~745--758.

\bibitem{Seo2000}
{\sc S.~Seo, M.~Wallat, T.~Graepel, and K.~Obermayer}, {\em Gaussian process
  regression: active data selection and test point rejection}, in Proceedings
  of the International Joint Conference on Neural Networks (IJCNN), vol.~3,
  IEEE, 2000, pp.~241--246.

\bibitem{Textor2009}
{\sc C.~Textor, H.~Graf, A.~Longo, A.~Neri, T.~E. Ongaro, P.~Papale,
  C.~Timmreck, and G.~G.~J. Ernst}, {\em Numerical simulation of explosive
  volcanic eruptions from the conduit flow to global atmospheric scales}, Ann.
  of Geophys., 48 (2009), pp.~817--842.

\bibitem{Van2009}
{\sc A.~W. van~der Vaart and H.~J. van Zanten}, {\em {Adaptive Bayesian
  estimation using a Gaussian random field with inverse gamma bandwidth}}, The
  Ann. of Statist., 37 (2009), pp.~2655--2675.

\end{thebibliography}


\begin{thebibliography}{1}

\bibitem{Gramacy2008}
Robert~B. Gramacy and Herbert K.~H. Lee.
\newblock {Bayesian treed Gaussian process models with an application to
  computer modeling}.
\newblock {\em J. Am. Statist. Assoc.}, 103:1119--1130, 2008.

\bibitem{Lopes2011}
H.~F. Lopes, C.~M. Carvalho, M.~S. Johannes, and N.~G. Polson.
\newblock {\em Particle learning for sequential Bayesian computation}.
\newblock Oxford University Press, 2011.

\end{thebibliography}

\end{document}



\maketitle
\vspace{.1cm}
\begin{center}{\large{\textbf{Web Appendix A}}}\\
\vspace{.2cm}
{\textbf{Two-dimensional numerical examples}}\end{center}
  \begin{figure}[h!]
  \centering
  \includegraphics[scale = .55, angle = -90]{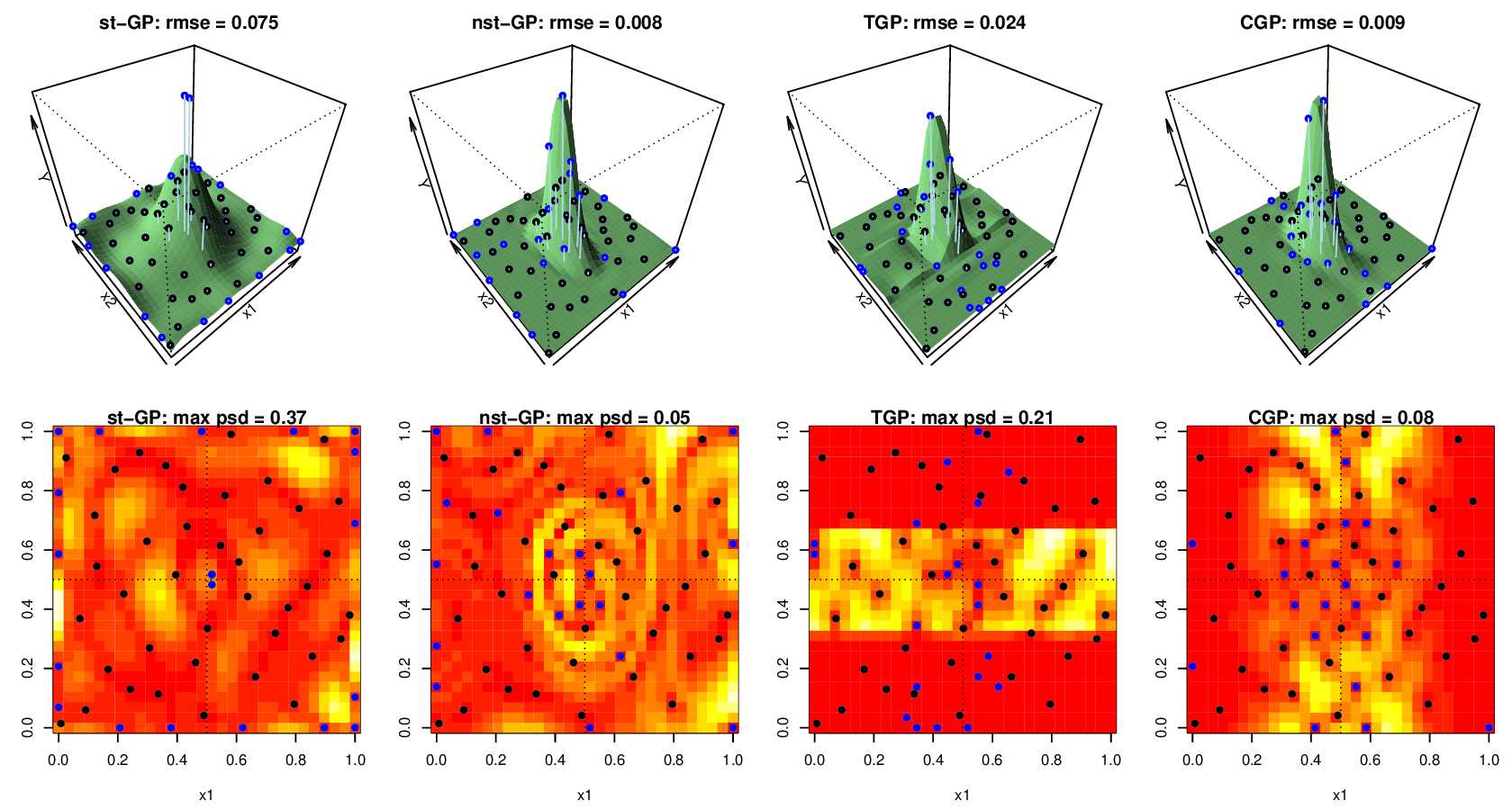}
  \caption*{{\textbf{Figure S1.}} Comparison between stationary GP (st-GP), non-stationary GP via latent input augmentation (nst-GP), TGP, and CGP on the 2D menhir function at $T = 60$. Black points: initial 40 LHD (common to all emulators); blue points: 20 additional points selected via ALM. Top row: posterior mean predictive surface, $\hat{f}$, and root mean squared error (rmse); bottom row: predictive standard deviation, $\hat{\sigma}$, and maximum predictive standard deviation (max psd). The quality of the prediction is assessed at a collection of 900 points in $\Omega = [0,1]^2$, i.e. an expanded grid of 30 equally spaced points along each coordinate axes. Note that the stationary GP does not interpolate at the peak. We defer a discussion on this phenomenon to Section 7 of the paper.}
   \label{nstat4}
\end{figure}

\newpage
  \begin{figure}[h!]
  \centering
   \includegraphics[scale = .55, angle = -90]{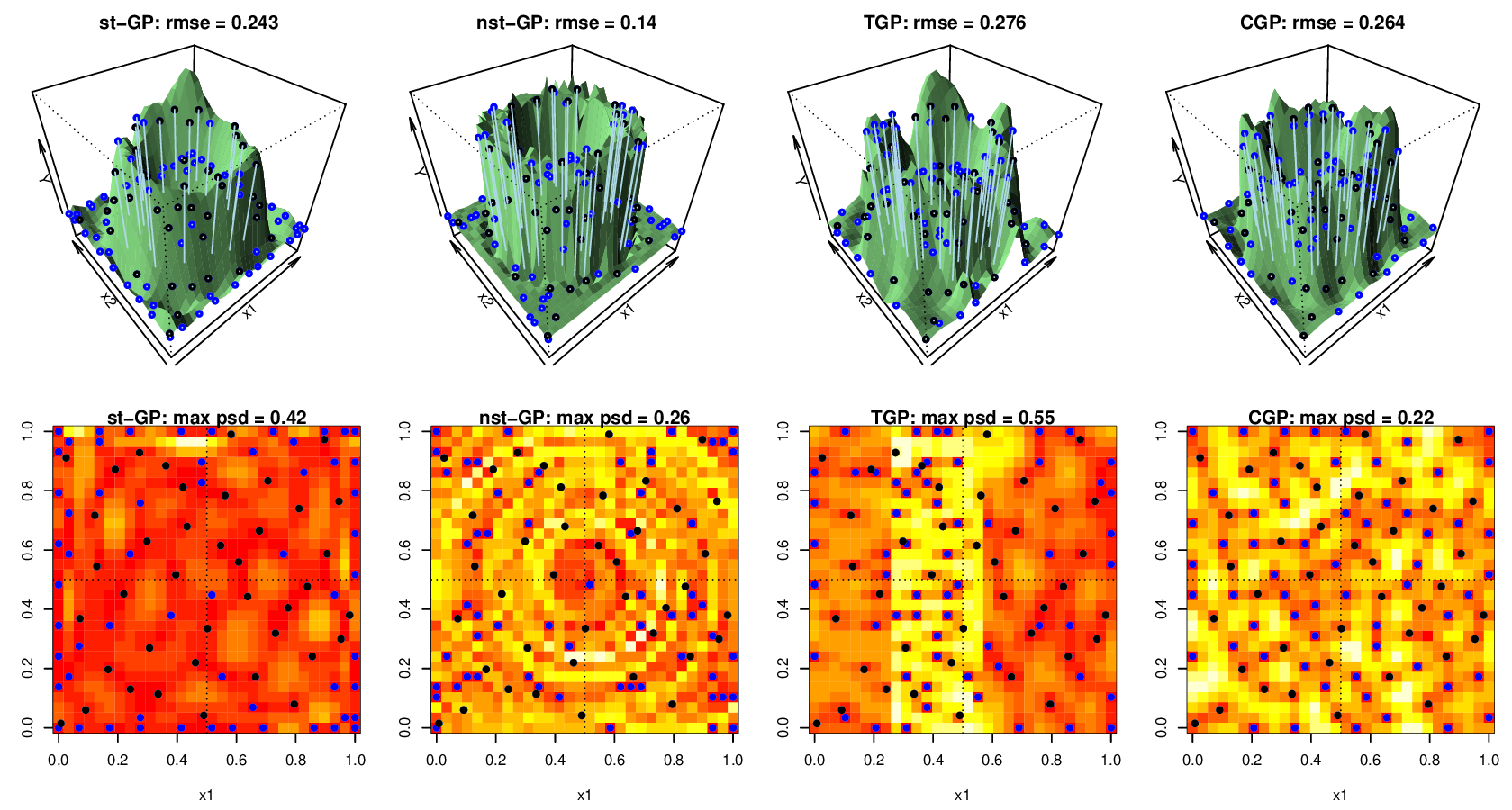}
  \caption*{{\textbf{Figure S2.}} Comparison between stationary GP (st-GP), non-stationary GP via latent input augmentation (nst-GP), TGP, and CGP on the 2D well function at $T = 100$. Black points: initial 40 LHD (common to all emulators); blue points: 60 additional points selected via ALM. Top row: posterior mean predictive surface, $\hat{f}$, and root mean squared error (rmse); bottom row: predictive standard deviation, $\hat{\sigma}$, and maximum predictive standard deviation (max psd). The quality of the prediction is assessed at a collection of 900 points in $\Omega = [0,1]^2$, i.e. an expanded grid of 30 equally spaced points along each coordinate axes.}
   \label{nstat6}
\end{figure}

\vspace{.1cm}
\begin{center}{\large{\textbf{Web Appendix B}}}\\
\vspace{.2cm}
{\textbf{Six-dimensional numerical examples}}\end{center}
Our methodology relies on two stationary GPs, one for the function of interest and one for the latent input: 
			\begin{equation}
			f \vert \bst \sim \text{GP}(\mu_{\bst}, C_{\bst}), \qquad \text{and} \qquad g \vert \bst \sim \text{GP}(0, \tilde{K}_{\bst}),
			\label{construction}
			\end{equation}
where $\bst$ is a vector of model parameters. To simplify the notations, we shall drop the $\bst$ subscript to the mean, covariance, and correlation functions hereafter. For our formulation, we choose
		\begin{align}
		&\mu(\bs) =  h(\bs)^\top \boldsymbol{\beta}, \quad \text{with} \quad h(\bs) = [1, \bs]^\top \\
		&C(\bs_i, \bs_j) = \sigma^2 K(\bs_i, \bs_j) \\ 
		&{K}(\bs_i, \bs_j) = \exp\left\{- \sum_{l = 1}^p \phi_l(x_{il} - x_{jl})^2 - \phi_{p+1}(Z_i - Z_j)^2\right\} \\
		&Z_i = g(\bs_i) \\
		& \tilde{K}(\bs_i, \bs_j) = \exp\left\{- \sum_{l = 1}^p \tilde{\phi}_l (x_{il} - x_{jl})^2\right\}.
		\label{end}
		\end{align}
Parameter $\phi_l \geq 0$ (or $\tilde{\phi}_l \geq 0$ for the latent correlation function $\tilde{K}$) controls the sensitivity of $f$ $(g)$ to $x_l$. For example, $\phi_l = 0$ $(\tilde{\phi}_l = 0)$ removes $x_l$ (dimension reduction), whereas larger $\phi_l$ $(\tilde{\phi}_l)$ denotes smaller correlation, i.e. $f(\bs)$ and $f(\bs^\prime)$ ($g(\bs)$ and $g(\bs^\prime)$) are less related in the $x_l$ direction and the function is more complex. \\
\indent We implement our emulator via particle learning (PL - \cite{Lopes2011}) as described in Section 3 of the paper. We train our emulator on a 120 LHD and then let it select 80 additional points via ALM, for a total of 200 points in the final design. Each particle contains an estimate of model parameters in (\ref{construction})-(\ref{end}). Figure S3 shows the distribution across particles of the estimated $\{\tilde{\phi}_i\}_{i = 1}^6$ for the 6D well example. We recall that true surface is
		\begin{equation}
	f(x_1, \dots, x_6) = 
		\left\{
		 \begin{array}{l l}
    			1, & \quad \text{if }\sum_{i = 1}^4 (x_i - 0.5)^2  > 0.025 \text{ and }  \sum_{i = 1}^4 (x_i - 0.5)^2  < 0.25\\
   			0, & \quad \text{otherwise}
			\label{nwell}
              \end{array} \right.	\end{equation}
on the hypercube $\boldsymbol{X} = [0,1]^6$. Therefore, $f$ in (\ref{nwell}) is constant in $x_5$ and $x_6$. Note that ${\tilde{\phi}}_5$ are ${\tilde{\phi}}_6$ are estimated to be smaller than $\{\tilde{\phi}_i\}_{i = 1}^4$, thus showing that our emulator is learning that $f$ is less sensitive to these input dimensions. Besides a few large isolated outliers, the correlation length parameters of $K$ are estimated to be small except for $\phi_7$, which is associated to the latent input $Z$ (Figure S4). This indicates that the latent input is needed to learn the surface well. At every input configuration $\bs$ corresponds an estimate of the latent input $Z = g(\bs)$. The last two panels show the estimated latent input at each design point, $g(\bs_1), \dots, g(\bs_{200})$, plotted versus the first and fourth dimension of the corresponding input configuration. Specifically, we plot $g(\bs_1)$ (with $\bs_1 = (x_{1,1}, \dots, x_{1,6})$) at $x_{1,1}$ (or $x_{1,4}$ in the last panel), $g(\bs_2)$ at $x_{2,1}$ (or $x_{2,4}$), and so forth. Despite being slightly elongated, the circular pattern of the well function emerges in the estimated latent input. 

  \begin{figure}[h!]
  \centering
 \includegraphics[scale = .62, angle = -90]{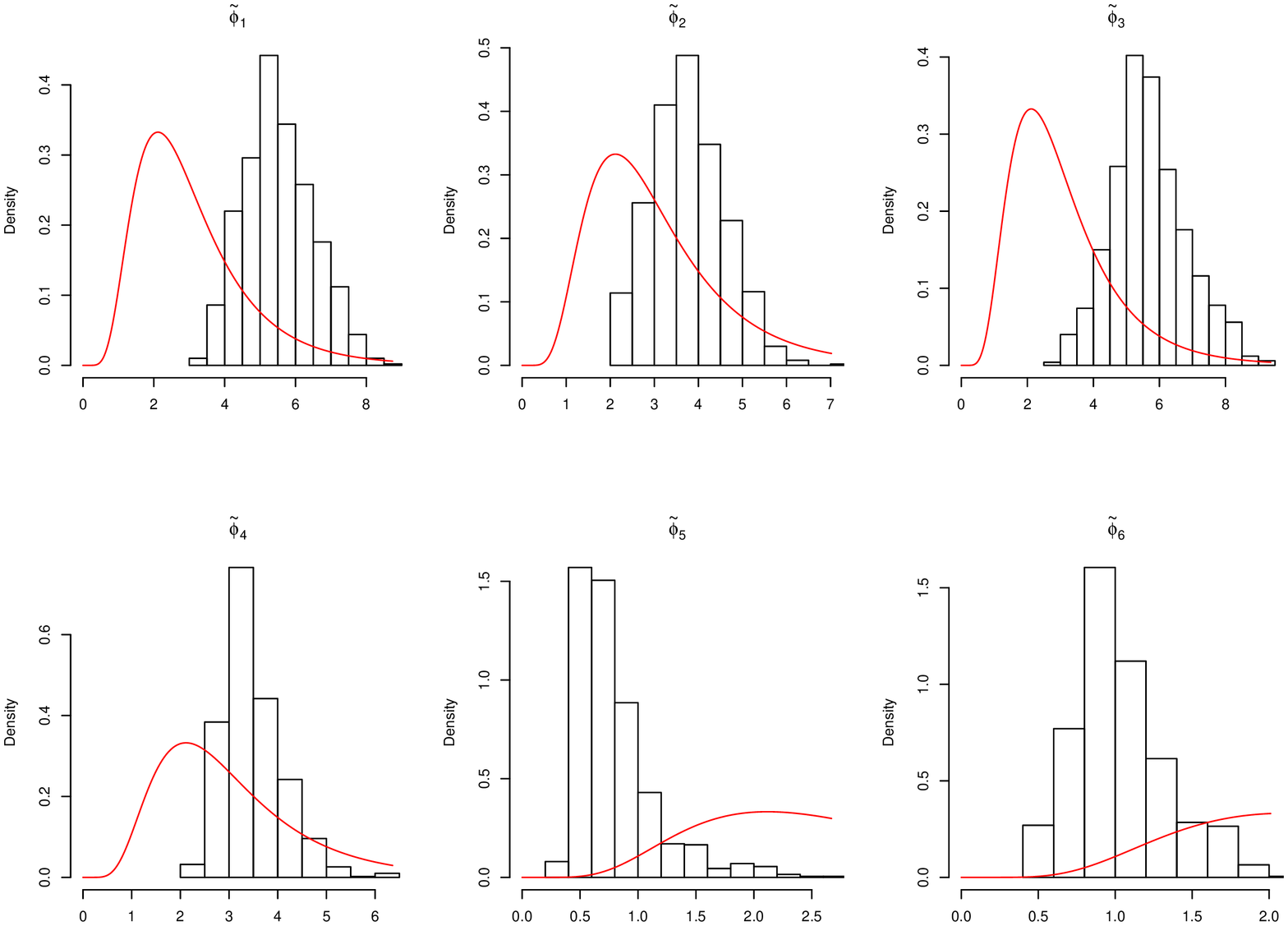}
  \caption*{{\textbf{Figure S3.}} Distribution across particles of the correlation length parameters of $\tilde{K}$, $\{\tilde{\phi}_i\}_{i = 1}^6$, at $T = 200$ (120 fixed LHD + 80 additional points selected via ALM) in the 6-dimensional well example (\ref{nwell}). The red curve denotes the prior distribution on the latent correlation parameters. Specifically, $\log \tilde{\phi}_i \sim \text{N}(0.5, 0.25), i = 1, \dots, 6$.}
   \label{6dwellphit}
\end{figure}

  \begin{figure}[h!]
  \centering
 \includegraphics[scale = .80, angle = -90]{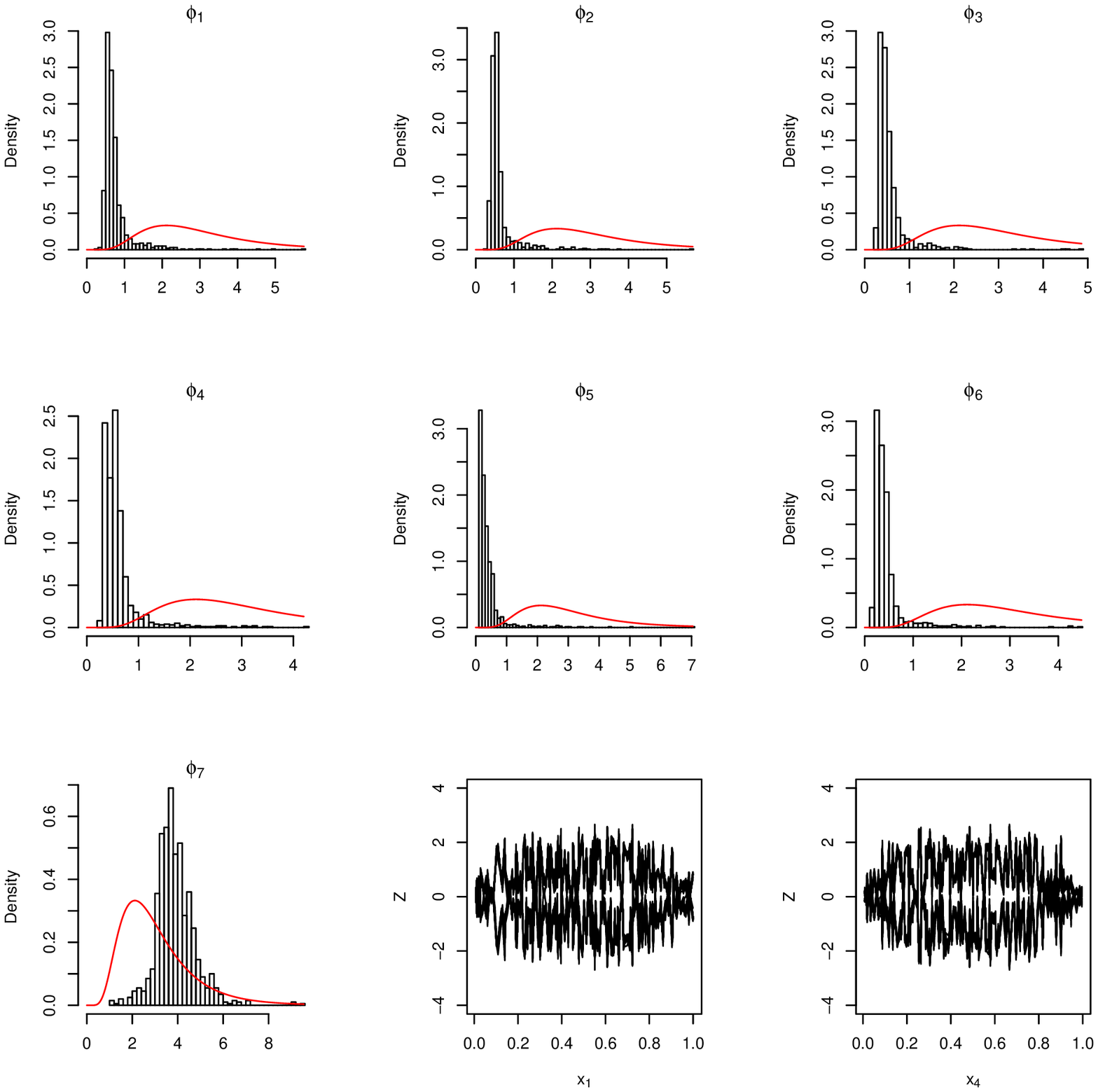}
 \hspace{-1cm}
  \caption*{{\textbf{Figure S4.}} Panels $\{\phi_i\}_{i = 1}^7$: distribution across particles of the correlation length parameters of ${K}$, $\{\phi_i\}_{i = 1}^7$,  at $T = 200$ (120 fixed LHD + 80 additional points selected via ALM) in the 6-dimensional well example (\ref{nwell}). The red curve denotes the prior distribution on the correlation parameters. Specifically, $\log {\phi}_i \sim \text{N}(1, 0.25), i = 1, \dots, 7$. The last two panels show the latent input estimated at each design point, $g(\bs_1), \dots, g(\bs_{200})$, plotted versus the first and fourth dimension of the corresponding input configuration. Specifically, plots show $g(\bs_1)$ at $x_{1,1}$ (or $x_{1,4}$ in the last panel), $g(\bs_2)$ at $x_{2,1}$ (or $x_{2,4}$), and so forth. }
   \label{6dwellphi}
\end{figure}

\clearpage
\begin{center}{\large{\textbf{Web Appendix C}}}\\
\vspace{.1cm}
{\textbf{LGBB CFD experiment}}\end{center}
\vspace{.2cm}
 \begin{figure}[hb!]
 \vspace{-0.8cm}
  \centering
 \includegraphics[scale = .65, angle = -90]{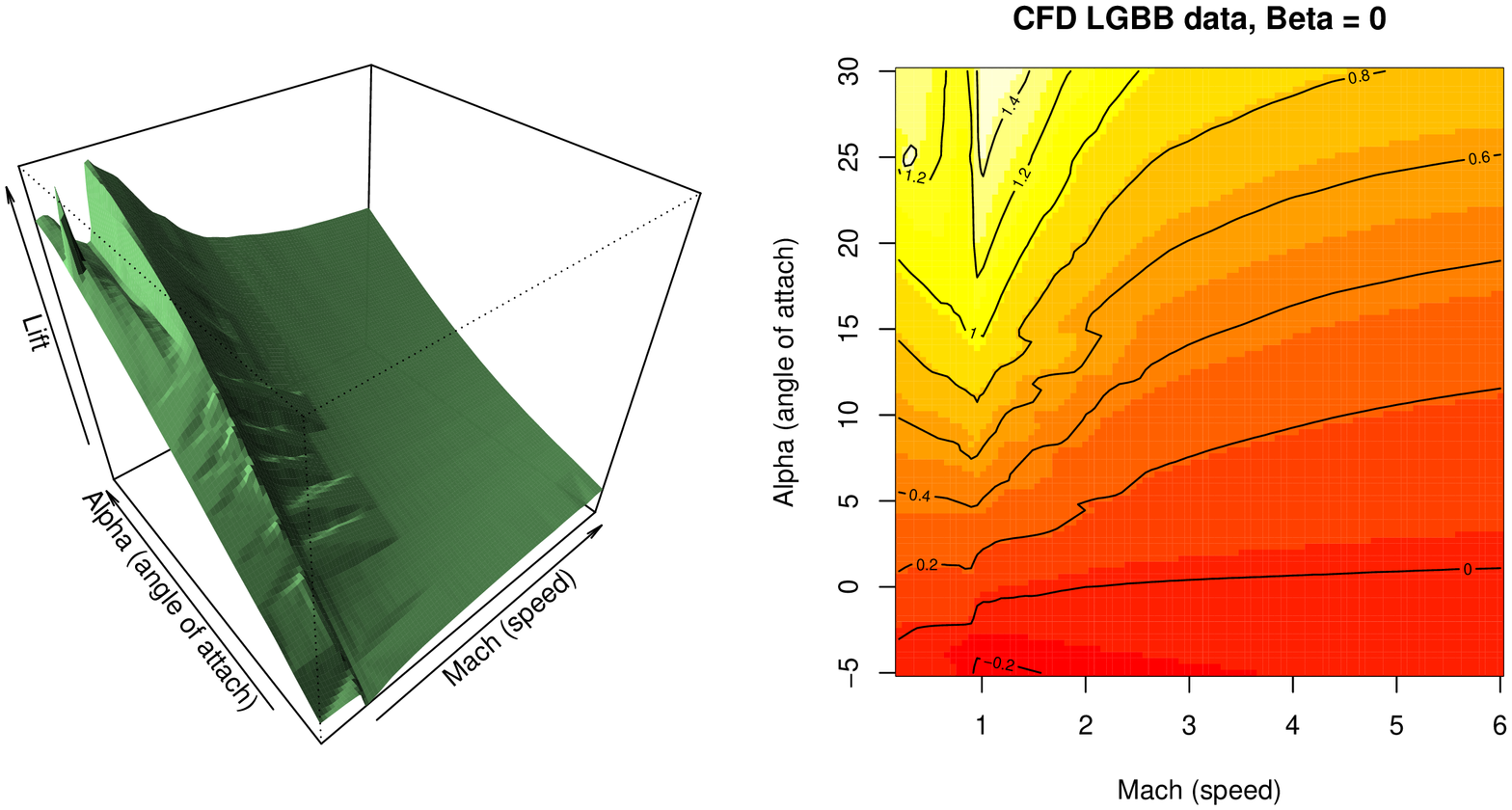}
 \vspace{-0.5cm}
  \caption*{{\textbf{Figure S5.}} Interpolated lift surface plotted as a function of Mach (speed) and Alpha (angle of attack) with Beta (side-slip angle) fixed to zero. The ridge at Mach 1 denotes a distinction between subsonic flows and supersonic flows. The upper-left corner of the plot (high angle of attack, low speed) shows a spike which is a result of false convergence of the simulator \cite{Gramacy2008}. }
  \end{figure}

 \clearpage
\vspace{.1cm}
\begin{center}{\large{\textbf{Web Appendix D}}}\\
\vspace{.2cm}
{\textbf{Two-stage empirical Bayes approximation of non-stationary GP emulator}}\end{center}
The particle learning (PL) implementation of our emulator is formulated around an effort at joint modelling of both $f$ and $Z$ (Section 3). The latent input $Z = g(\bs)$ essentially becomes a vector of model parameters that can not be marginalised out within our construction, thus needs to be learnt. The learning of the latent input vector, which is of sequentially increasing dimension within our PL construction, can face challenges and a large amount of data may be needed to learn it well. In alternative to the PL ``full Bayes'' approach, we elect here a cruder but much simpler strategy to approximate our non-stationary emulator. Specifically, we can rely on a two-stage approach where the first stage focuses on the estimation of the latent predictor, and the second stage focuses on learning $f$ assuming that the latent predictor is known and fixed at the level estimated in the first stage. Although several methods exist to obtain an estimate of the latent input GP at first stage, we investigate here an MCMC-based nonparametric regression approach. \\
\indent Suppose we are given an initial design $\{\boldsymbol{x}_i, f(\bs_i)\}_{i = 1}^t$. We consider $Z = g(\bs)$ and model $g$ by a stationary GP indexed by the $p$-dimensional vector of known inputs, $\bs$, and a vector of model parameters, $\bst$:
	\begin{equation}
	g\vert \bst \sim \text{GP}(\tilde{\mu}_{\bst}, \tilde{K}_{\bst}).
	\label{g}
	\end{equation}	
\indent Several choices are available for the GP mean $\tilde{\mu}_{\bst}$ including the constant-zero mean. The correlation function $\tilde{K}_{\bst}$ corresponds to Equation (2.8) in Section  2.2. 
If we consider a unit scale and fix $\tilde{\mu}_{\bst} \equiv 0$, (\ref{g}) reduces to the GP prior on $g$ as of Expression (2.7) in Section 2.2.
	One can estimate $g$ from a smooth GP (noisy) regression:
		\begin{equation} 
		f(\bs_i) = g(\boldsymbol{x_i}) + \epsilon_i,  \quad \text{with} \quad \epsilon_i \sim \text{N}(0, \tau^2), \quad \text{and} \quad i = 1, \dots, t.
		\label{latent}
		\end{equation}
\noindent Overall, the model in (\ref{g})-(\ref{latent}) is equivalent to assuming a GP prior on $f$:
 	\begin{equation} 
 	f \vert \bst, \tau^2 \sim \text{GP}(\tilde{\mu}_{\bst}, \tilde{K}_{\bst }+ \tau^2\delta_{i,j}),
	\label{f}
	\end{equation} 
where $\delta_{\cdot, \cdot}$ is the Kronecker delta function. Bayesian inference of (\ref{f}) proceeds via MCMC by drawing realisations from the joint posterior distribution of the model parameters. For all $\bs$ of interest and using the parameter values drawn from the joint posterior at iteration $k$, we can estimate $Z$ at $\bs$ as $\hat{g}_k({\bs}) = \mathbb{E}[f(\bs) \vert \bs, \{\bst, \tau^2\}_k] $, the point predictor of $f$ at $\bs$ at iteration $k$. Steps are repeated a large number of times, and the average of the point predictors is used as estimate of $g$. At the second stage, we consider $f \vert \hat{g}, \bst \sim \text{GP}(\mu_{\bst}, K_{\bst})$, where $f$ is now a stationary GP indexed by a $p+1-$dimensional vector of inputs $\{\bs, \hat{g}(\bs)\}$ as in (2.5)-(2.6) (Section 2.2) 
under the fiction that the latent input is known. Inference for $\boldsymbol{\theta}$ proceeds via MCMC and sequential design is embedded to guide the selection of new inputs. \\ 
\indent The two-stage, MCMC-based inference is perfectly coherent and comes closest to a full Bayesian treatment of the problem in that it takes into account uncertainty in estimating the hyperparameters and the latent input GP at first stage. 
However, MCMC-based inference is ill-suited to sequential design, as the chain must be restarted and iterated until convergence when the design is augmented with a new pair [$\bs_{t+1}, f(\bs_{t+1})$]. Fits from previous iterations can only guide the i\-ni\-ti\-a\-li\-za\-tion of the new Markov Chain. 

\begin{center}{\textbf{Simulation studies}}\end{center}

We evaluate the performance of the two-stage approximation on the sequential experiments presented in Section 5.  \\
\indent Figure S6 shows the performance of the two-stage approximation on the 2D examples of Section 5.1 given the initial 40 LHD (black points). Figure S7 reports the predictive surface and standard deviation after 20 additional inputs (blue points) are selected based on uncertainty. The patterns observed in the predictive standard deviation (panels in the second row of Figure S6 and S7) show that the emulator is learning the geometry of the local features. The emulator creates a balance between an {\textit{exploration}} of the input space via the LHD (given design) and the ({\textit{exploitation}}) of the local features as shown by the higher concentration of new inputs at the edges of the features. \\
\indent Figure S8 shows a comparison between two-stage approximation and PL ``full Bayes"  in terms of progression of the root mean square error (RMSE). Top panels refer to the 2D examples outlined in Section 5.1: no implementation is preferred in terms of predictive accuracy across functions or number of input points. The ``full Bayes'' approach is preferred on the well function (top right), whereas two-stage is preferred on the Menhir function for small designs (top left). This is probably due to the ability of the two-stage approximation in learning better the latent input GP through the first noisy regression, which results into quicker learning of $f$. However, the predictive accuracy of the ``full Bayes'' approximation of our emulator considerably improves when one point is selected at the center of the input space (this happens at $t=48$), and eventually reconciles with two-stage approximation. Bottom panels in Figure S8 show that the full version of our emulator outperforms the two-stage approximation on the 6D and NASA experiments.\\
\indent To conclude, the two-stage approximation of our emulator preserves some good features of the ``full Bayes'' version, namely learning the geometry of different types of shape and increasing the sampling frequency of new inputs along important input dimensions. Therefore, it constitutes a valid alternative to the full Bayesian implementation for adaptive design selection and function approximation. However, the ``full Bayes'' version often achieves lower RMSE, in particular for larger designs or in higher dimensions.
\begin{figure}[h]
 \centering
\includegraphics[scale = .57, angle = -90]{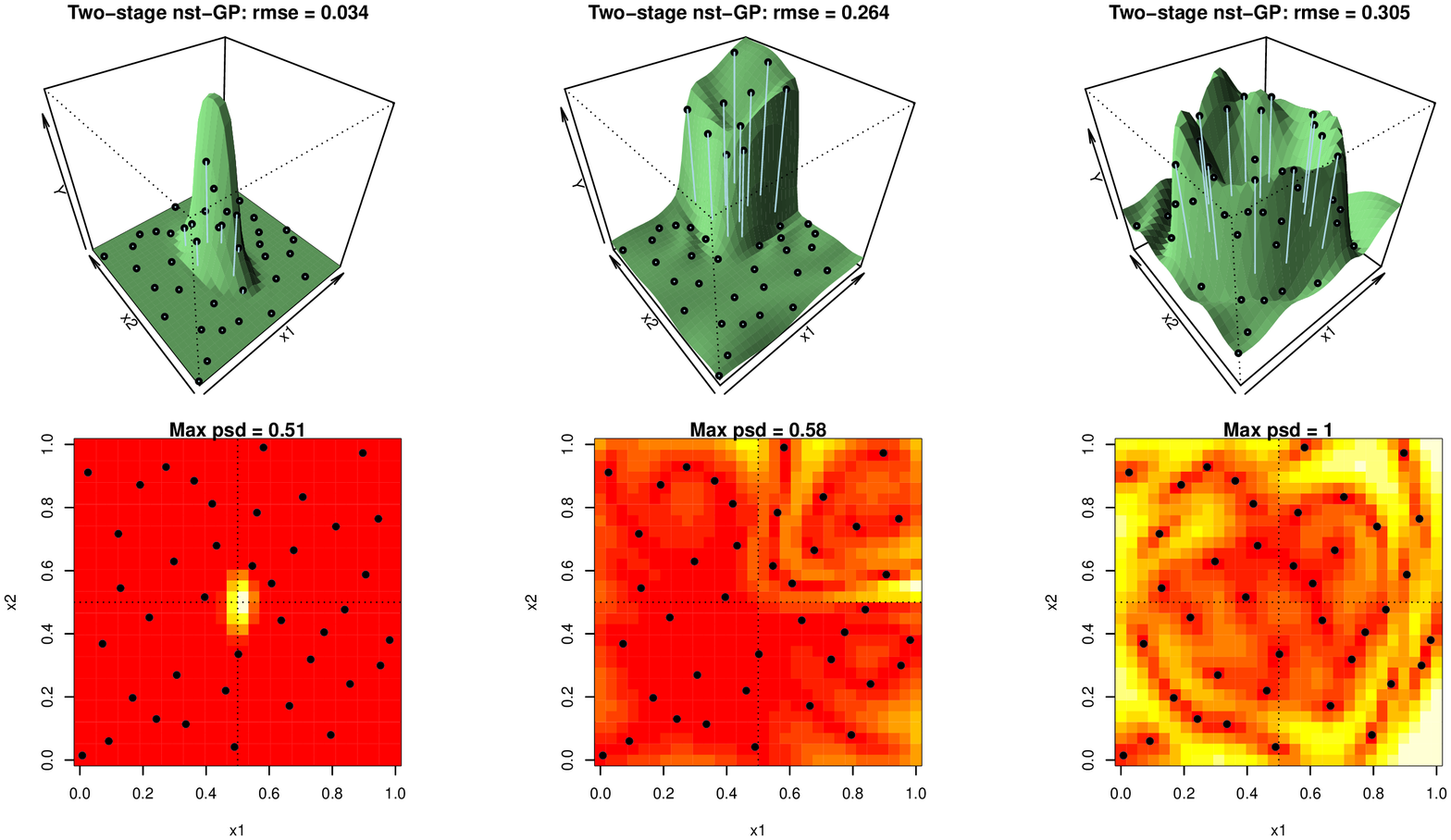}
\caption*{{\textbf{Figure S6.}} Predictive surface and standard deviation at a set of 900 predictive points $[0,1]^2$ obtained with the two-stage MCMC-based implementation of the non-stationary GP emulator on the 2D Menhir, building, and well examples numerical examples. Quantitative summaries report the root mean squared error (RMSE) and the maximum predictive standard deviation (pred sd) computed based on the test points. The fit is based on the same 40 LHD (black points) that was used in our 2D numerical examples in Section 5.1.}
 \label{Surya2D}
\end{figure}
 \begin{figure}[h]
  \centering
 \includegraphics[scale = .57, angle = -90]{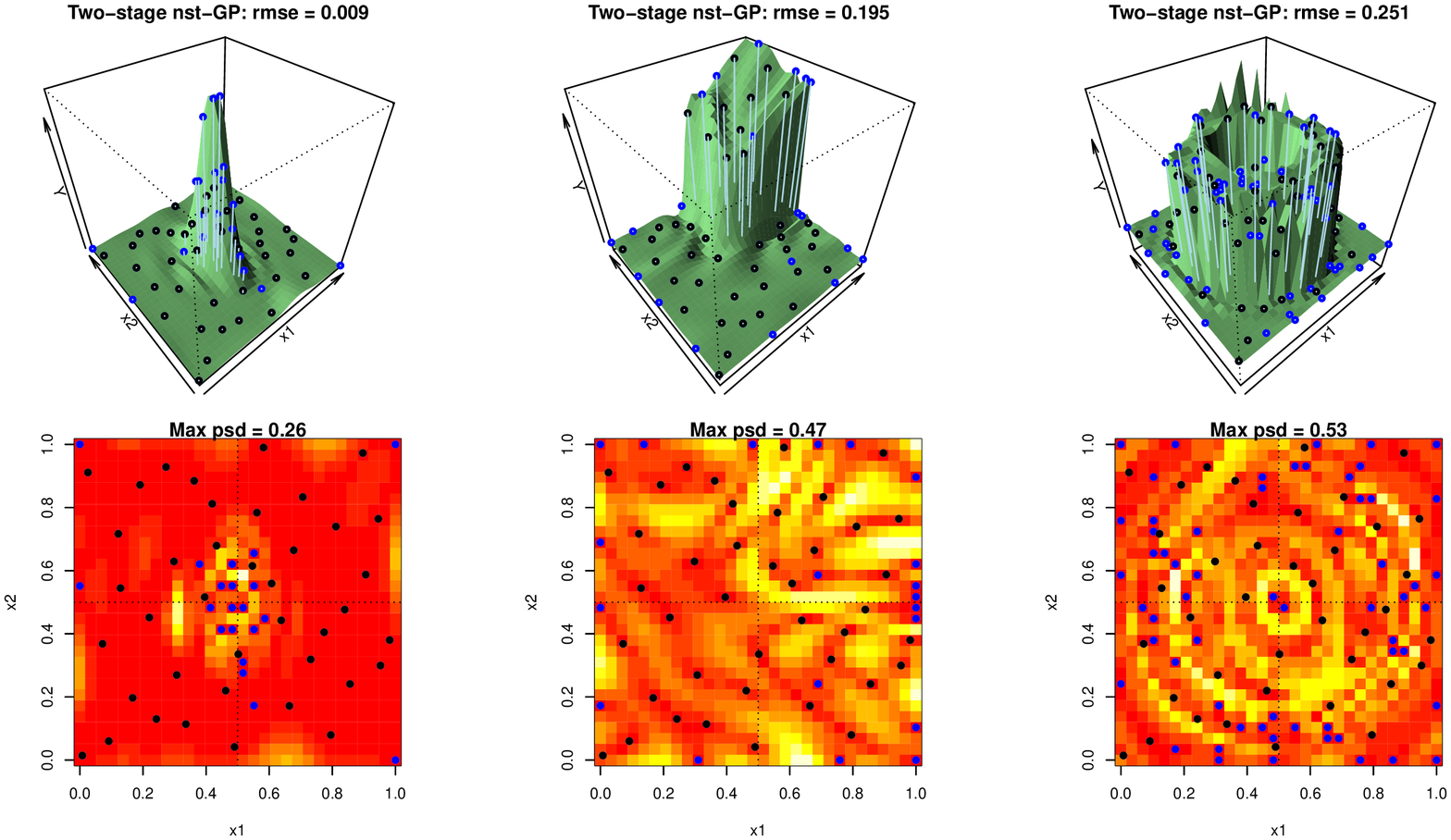}
 \caption*{{\textbf{Figure S7.}} Predictive surface and standard deviation at a set of 900 predictive points $[0,1]^2$ obtained with the two-stage MCMC-based implementation of the non-stationary GP emulator on the 2D Menhir, building, and well examples numerical examples. Quantitative summaries report the root mean squared error (RMSE) and the maximum predictive standard deviation (pred sd) computed based on the test points. Black points: initial 40 LHD (given). Blue points: additional points selected via ALM criterion.}
   \label{2DTwoStage}
\end{figure}  
 \begin{figure}[t!]
  \centering
 \includegraphics[scale = .60, angle = -90]{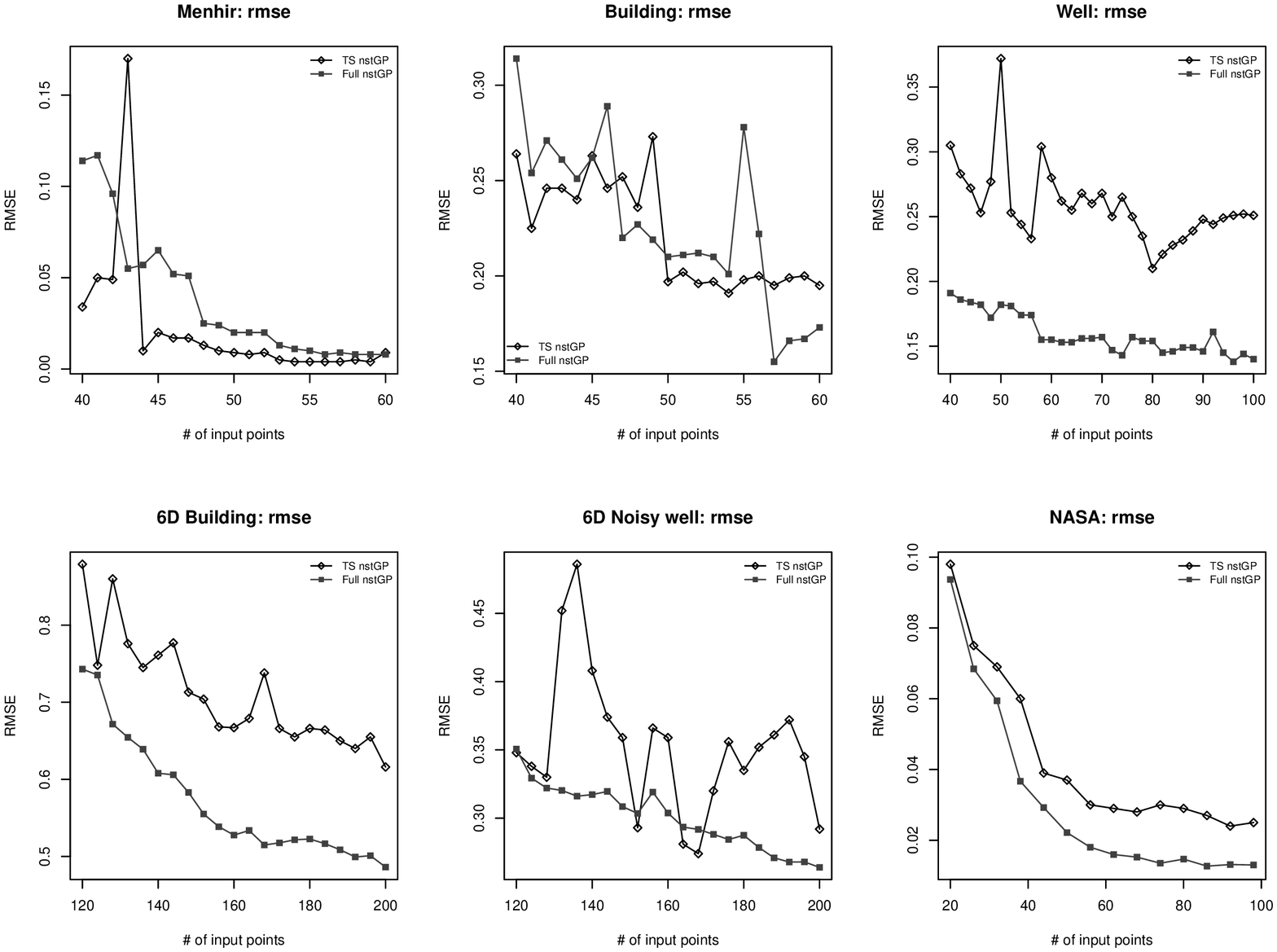}
 \caption*{{\textbf{Figure S8.}} Comparison between two-stage approximation (TS nstGP) and ``full Bayes'' (Full nstGP) in terms of RMSE progression as additional input points are being selected. Top panels: 2D Menhir, building and well functions. Bottom panels: 6D building and well examples and NASA rocket booster experiment.}
   \label{2DTwoStagermse}
\end{figure}


\bibliographystyle{plain}
\bibliography{samsi}